\documentclass[aip,jcp,reprint,amsmath,amssymb,floatfix,citeautoscript,nofootinbib]{revtex4-2}
\usepackage{times}
\usepackage{mathptmx}
\DeclareMathAlphabet{\mathcal}{OMS}{cmsy}{m}{n} 
\usepackage[T1]{fontenc}
\usepackage{algorithm}
\usepackage{algpseudocode}
\usepackage{bm}
\usepackage{amsmath}
\usepackage{amssymb}
\usepackage{xcolor}
\usepackage{graphicx}
\usepackage{hyperref}
\hypersetup{
    colorlinks=true,
    linkcolor=blue,
    urlcolor=blue,
    citecolor=blue,
}
\usepackage{cleveref}
\usepackage{braket}
\usepackage[version=4]{mhchem}
\usepackage{multirow}
\usepackage{booktabs}

\newcommand{\mi}{\mathrm{i}}
\newcommand{\me}{\mathrm{e}}

\definecolor{myred}{rgb}{0.8,0,0}
\newcommand{\myred}[1]{{#1}}

\begin{document}

    \title{Smoothed truncated Coulomb potential for periodic Gaussian-basis Hartree--Fock exchange}

    \author{Gengzhi Yang}
    \affiliation{Joint Center for Quantum Information and Computer Science, University of Maryland, College Park, Maryland, 20742}
    \affiliation{Department of Mathematics, University of Maryland, College Park, Maryland, 20742}

    \author{Aamy Bakry}
    \affiliation{Department of Chemistry and Biochemistry, University of Maryland, College Park, Maryland, 20742}

    \author{Hong-Zhou Ye}
    \email{hzye@umd.edu}
    \affiliation{Department of Chemistry and Biochemistry, University of Maryland, College Park, Maryland, 20742}
    \affiliation{Institute for Physical Science and Technology, University of Maryland, College Park, Maryland, 20742}

    \date{\today}

    \begin{abstract}
        Truncated Coulomb (TC) potentials reduce finite-size errors and accelerate thermodynamic-limit convergence in periodic Hartree--Fock (HF) calculations, but their use with Gaussian basis sets is complicated by the evaluation of electron repulsion integrals (ERIs), particularly for nonspherical truncation domains and all-electron calculations.
        We introduce the smoothed truncated Coulomb (sTC) potential as a systematically improvable approximation to a parent TC potential.
        A real-space Gaussian convolution smooths the sharp truncation boundary, and a single dimensionless parameter, $\eta$, controls the width of the smoothing window, which can be tightened to systematically approach the TC reference.
        The smoothing by sTC enables a dual-space algorithm for evaluating periodic Gaussian-basis ERIs that requires neither a large plane-wave basis nor new molecular integral kernels and is applicable to both pseudopotential and all-electron calculations, including the important Wigner--Seitz-cell truncation boundaries.
        Benchmarks spanning insulating, semiconducting, layered, metallic, and molecular-crystal systems show that sTC-based HF closely reproduces TC results in pseudopotential calculations and extends TC-quality calculations to all-electron settings.
        Across these systems, sTC substantially improves thermodynamic-limit convergence relative to the probe-charge Ewald method while remaining practical for all-electron calculations in which direct TC-based calculations are computationally challenging.
    \end{abstract}

    \maketitle

    \section{Introduction}

    The Hartree--Fock (HF) exchange energy $E_{\text{HFX}}$ is central to both HF theory~\cite{Szabo96Book} and Kohn--Sham density functional theory (KS-DFT)~\cite{Kohn65PR} with hybrid exchange--correlation functionals~\cite{Becke93JCP,Perdew96JCP}.
    The resulting mean-field states also provide reference wave functions for post-HF correlation and excited-state methods~\cite{Szabo96Book,Bartlett2007,Booth13Nature,Schirmer25PCCP}.
    Under periodic boundary conditions, however, evaluating $E_{\text{HFX}}$ on a finite $k$-point mesh requires careful treatment of the singular contribution associated with the long-range Coulomb interaction~\cite{Gygi86PRB,Paier05JCP,Spence08PRB,Broqvist09PRB,Sundararaman13PRB,Ye26arXiv}.
    Without such treatment, the singularity produces slowly decaying finite-size errors (FSEs) and impedes convergence to the thermodynamic limit (TDL).

    Several reciprocal-space strategies have been developed to treat the HF exchange singularity and reduce the associated FSE.
    In reciprocal space, the Coulomb potential has an $\mathcal{O}(q^{-2})$ singularity near $q = 0$,
    \begin{equation}    \label{eq:v_q_coulomb}
        v(\bm{q})
            = 4\pi/q^2,
    \end{equation}
    where $\bm{q}$ is a reciprocal-space coordinate and $q = |\bm{q}|$.
    Simply omitting the $q=0$ component leads to an impractically slow $\mathcal{O}(N_k^{-1/3})$ decay of the FSE with the number of $k$-points sampling the first Brillouin zone.
    The auxiliary-function method~\cite{Gygi86PRB,Wenzien95PRB} and the closely related probe-charge Ewald method~\cite{Paier05JCP,Broqvist09PRB} improve this scaling to $\mathcal{O}(N_k^{-1})$---also known as inverse-volume scaling~\cite{Xing24PRX}---by modifying the $q=0$ component to remove the leading-order error.
    Incorporating curvature information near $q=0$ into the auxiliary function can further improve the scaling to $\mathcal{O}(N_k^{-2})$~\cite{Carrier07PRB}, and optimized auxiliary functions have been explored within singularity-subtraction schemes to obtain still more rapid FSE decay~\cite{Quiton25JCTC,Quiton26arXiv}.
    An alternative is the staggered-mesh method~\cite{Xing21JCTC,Quiton24JCTC}, which avoids the Coulomb divergence by evaluating exchange between two mutually shifted $k$-point meshes \myred{and improves the FSE scaling to $\mathcal{O}(N_k^{-5/3})$ in gapped solids.}

    From a real-space perspective, the FSE arises from the periodic repetition of the one-particle density matrix, which introduces spurious exchange interactions between periodic images through the long-range Coulomb potential~\cite{Pisani80IJQC,Guidon09JCTC,Irmler18JCTC,Wang20JCP,Spence08PRB,Sundararaman13PRB}.
    The truncated Coulomb (TC) method removes these interactions by truncating the Coulomb potential at a finite boundary,
    \begin{equation}
        v_{\text{TC}}(\bm{r})
            = \frac{\Theta(\bm{r})}{r},
    \end{equation}
    where $\Theta(\bm{r})=1$ inside the truncation domain and vanishes otherwise.
    The use of TC for evaluating $E_{\text{HFX}}$ was first explored by Spencer and Alavi~\cite{Spence08PRB} using spherical truncation and was later generalized by Sundararaman and Arias~\cite{Sundararaman13PRB} to the Wigner--Seitz (WS) cell of the Born--von Karman (BvK) supercell.
    We refer to these two schemes as Sph-TC and WS-TC, respectively.
    For gapped systems and finite-temperature metals, WS-TC yields exponential FSE decay in $E_{\text{HFX}}$ and is typically more accurate than other FSE corrections at a given $k$-point mesh or supercell~\cite{Sundararaman13PRB}.
    TC has since been adapted to improve the finite-size convergence of other theories containing HF exchange or exchange-like contributions~\cite{Ren21PRM,Arruabarrena23PRB,Zhang26JCTC}.

    The original TC formulations and implementations were developed in a plane-wave framework~\cite{Spence08PRB,Sundararaman13PRB}, in which the TC kernel can be evaluated efficiently in reciprocal space.
    With Gaussian basis sets, the central challenge is instead the evaluation of electron repulsion integrals (ERIs) over the TC kernel.
    For Sph-TC, Guidon and co-workers~\cite{Guidon09JCTC} combined conventional recurrence relations for Gaussian ERIs~\cite{Ahlrichs06PCCP} with a sophisticated numerical scheme for evaluating the modified Boys functions required by the truncated potential.
    The resulting periodic HF implementation in CP2K~\cite{Kuhne20JCP,Iannuzzi26JPCB} has enabled efficient Gaussian-basis hybrid DFT calculations of periodic systems~\cite{Bussy24JCP}.
    Irmler and co-workers~\cite{Irmler18JCTC} subsequently introduced a simple algebraic evaluation of the modified Boys functions, leading to a robust Gaussian-basis implementation of Sph-TC-based HF exchange in TURBOMOLE~\cite{Furche14WIRCMS,Balasubramani20JCP}.
    Neither approach, however, has been generalized to the WS-TC potential.

    \myred{The Gaussian-basis HF implementation in PySCF~\cite{Sun18WIRCMS,Sun20JCP,Sun26JCP} supports WS-TC-based HF exchange.}
    In this implementation, the required TC-ERIs are evaluated in reciprocal space by expanding the Gaussian pair densities in an auxiliary plane-wave basis, an approach termed fast Fourier transform density fitting~\cite{Sun17JCP} (FFTDF) in PySCF and closely related to the Gaussian and plane-wave (GPW) method in CP2K~\cite{VandeVondele05CPC} \myred{and the QC-PBC module~\cite{Chen25JPCL} of Q-Chem~\cite{Shao15MP}.
    (In principle, implementing WS-TC in CP2K, Q-Chem, and other codes employing FFTDF or GPW should be technically straightforward.)}
    Because resolving the strongly localized core contributions in all-electron pair densities requires an impractically large plane-wave basis, this FFTDF-based approach is effectively limited to pseudopotential calculations.
    A similar limitation arises in other local-orbital representations.
    For example, the numerical-atomic-orbital implementation of HF exchange in FHI-aims~\cite{Blum09CPC} uses a spherical cut-Coulomb potential~\cite{Ren21PRM,Zhang26JCTC} adapted from the Spencer--Alavi Sph-TC scheme, but an extension to the WS-TC kernel has not been reported.
    An efficient local-orbital implementation of WS truncation that remains practical for all-electron calculations is therefore lacking.

    In this work, we introduce a \textit{smoothed truncated Coulomb} (sTC) potential that enables WS-TC-quality HF exchange calculations with Gaussian basis sets for both pseudopotential and all-electron treatments.
    The key idea is to smooth the parent TC kernel so that its long-range component decays rapidly in reciprocal space and can be represented using a modest plane-wave basis, while its short-range component can be evaluated through a rapidly convergent real-space lattice sum.
    The construction applies to general truncation domains for which the reciprocal-space TC kernel can be constructed, with the practically important WS boundary serving as the focus of the present work.
    The short-range lattice sum requires only ERIs over the complementary-error-function-attenuated Coulomb potential~\cite{Adamson97JMS},
    which are available in most Gaussian integral libraries~\cite{Sun15JCC,Libint2} because of their widespread use in range-separated hybrid DFT~\cite{Iikura01JCP,Heyd03JCP,Yanai04CPL} and wave-function theories~\cite{Angyan05PRA,Goll05PCCP,Kalai19JCP}.

    We implement two complementary algorithms for sTC-based periodic HF in PySCF.
    A direct reciprocal-space implementation based on FFTDF provides controlled comparisons between sTC and TC in pseudopotential calculations.
    The main implementation advance is a dual-space algorithm based on range-separated density fitting (RSDF)~\cite{Ye21JCPa,Ye21JCPb}, which exploits the short- and long-range decomposition above and avoids resolving compact core densities in the plane-wave basis.
    The resulting RSDF implementation is applicable to both pseudopotential and all-electron calculations.
    In pseudopotential calculations, benchmark results show that sTC closely reproduces WS-TC exchange energies, band structures, and structural properties over a practical range of the smoothing parameter.
    With the same smoothing parameters, all-electron sTC calculations for both gapped and metallic solids exhibit FSE trends parallel to their pseudopotential counterparts and yield consistently better FSE scaling than the probe-charge Ewald method.

    The remainder of this article is organized as follows.
    \Cref{sec:theory} defines the sTC potential as a systematically improvable approximation to a given TC potential, introduces the dimensionless inverse relative smoothing width $\eta$ as the main knob for accuracy control, and develops the dual-space RSDF formulation.
    \Cref{sec:comp_details} provides the computational details, and \cref{sec:results} presents benchmarks for gapped and metallic systems that establish $\eta=3$ for gapped systems and $\eta=4$ for metals as practical choices for closely reproducing WS-TC results.
    \Cref{sec:conclusion} summarizes the main findings and discusses future directions.

    \section{Theory}
    \label{sec:theory}

    \subsection{The smoothed truncated Coulomb (sTC) potential}
    \label{subsec:stc}

    For a given TC potential, we define the corresponding sTC potential as
    \begin{equation}    \label{eq:v_stc_def}
        v_{\text{sTC}}(\bm{r}; \omega)
            = \frac{1}{r}
            - \frac{\omega^3}{\pi^{3/2}}\int
            \mathrm{d}\bm{r}'\,
            \left[
                \frac{1}{r'}
                - v_{\text{TC}}(\bm{r}')
            \right] \mathrm{e}^{-\omega^2 |\bm{r}-\bm{r}'|^2}.
    \end{equation}
    This construction convolves the complementary part of the TC potential with a normalized Gaussian and subtracts the result from the full Coulomb potential.
    In real space, the Gaussian convolution smooths the TC potential near the truncation boundary over a characteristic length scale of $\mathcal{O}(\omega^{-1})$, rendering the sTC potential infinitely differentiable away from the Coulomb singularity at $r = 0$.
    The parameter $\omega$ therefore controls the fidelity of the sTC potential to its parent TC potential, with the original TC potential recovered in the limit $\omega \rightarrow \infty$.

    \begin{figure}[!t]
        \centering
        \includegraphics[width=3.0in]{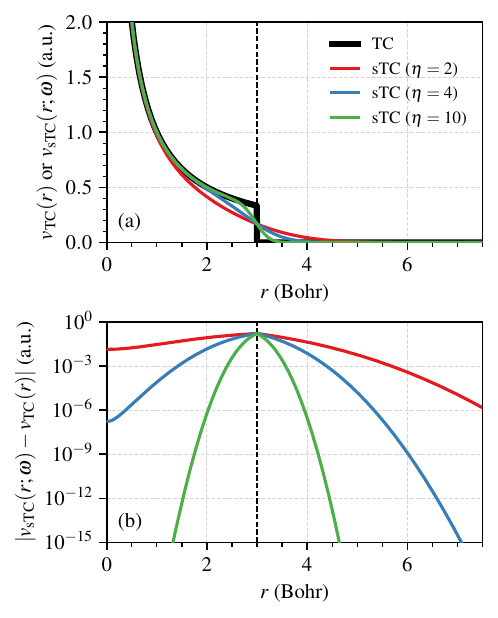}
        \caption{Comparison of the Sph-sTC potential for several values of $\eta=\omega R_{\text{c}}$ with the Sph-TC potential.
        $R_{\text{c}} = 3$~Bohr is used for both potentials.
        Panel (a) shows the potentials and panel (b) shows the absolute differences between the Sph-sTC potentials and the Sph-TC potential.
        }
        \label{fig:kernel}
    \end{figure}

    This behavior is illustrated explicitly for spherical truncation in \cref{fig:kernel}.
    For the Sph-TC potential~\cite{Spence08PRB} with a cutoff radius $R_{\text{c}}$,
    \begin{equation}
        v_{\text{Sph-TC}}(r; R_{\text{c}})
            = \begin{cases}
                \dfrac{1}{r}, & r \leq R_{\text{c}}, \\
                0, & r > R_{\text{c}},
            \end{cases}
    \end{equation}
    the corresponding Sph-sTC potential has the compact form
    \begin{equation}    \label{eq:v_sph_stc_r}
    \begin{split}
        v_{\text{Sph-sTC}}(r; \omega, R_{\text{c}})
            &= \frac{
                \text{erfc}(\omega(r-R_{\text{c}}))
                + \text{erfc}(\omega(r+R_{\text{c}}))
            }{2r}.
    \end{split}
    \end{equation}
    The dimensionless product $\omega R_{\text{c}}$ characterizes the width of the smooth region relative to the cutoff radius.
    Let $x = r/R_{\text{c}}$ denote the relative radial coordinate.
    Away from the truncation boundary, such that $\omega R_{\text{c}} |x-1| \gg 1$, the Sph-sTC potential converges exponentially to the Sph-TC potential,
    \begin{equation}    \label{eq:v_sph_stc_r_error}
        |v_{\text{Sph-sTC}}(r; \omega, R_{\text{c}})
        - v_{\text{Sph-TC}}(r; R_{\text{c}})|
            \sim \frac{ \omega\,
                \mathrm{e}^{-(\omega R_{\text{c}})^2 (x - 1)^2}
            }{
                2\sqrt{\pi} (\omega R_{\text{c}})^2 x |x - 1|
            },
    \end{equation}
    with the exponential convergence rate controlled by $\omega R_{\text{c}}$.
    Near the boundary, the Sph-sTC potential smoothly transitions from $1/r$ to zero as $r$ increases.
    At the boundary, the Sph-sTC potential rapidly approaches the midpoint value $1/(2R_{\text{c}})$ as $\omega R_{\text{c}}$ increases,
    \begin{equation}
        v_{\text{Sph-sTC}}(R_{\text{c}}; \omega, R_{\text{c}})
            \sim
            \frac{1}{2 R_{\text{c}}} +
            \frac{
                \omega\, \me^{-(2 \omega R_{\text{c}})^2}
            }{
                \sqrt{\pi} (2 \omega R_{\text{c}})^2
            },
        \quad{}(\omega R_{\text{c}} \gg 1).
    \end{equation}

    For a general truncation domain, a compact analytical form of the sTC potential is typically unavailable.
    Nevertheless, the spherical analysis provides a useful qualitative picture: the Gaussian convolution smooths the TC potential within a boundary layer of characteristic width $\mathcal{O}(\omega^{-1})$, while the sTC potential rapidly approaches the parent TC potential away from this layer.
    For a truncation boundary represented by a direction-dependent radial cutoff $R(\theta,\phi)$, the local relative smoothing width is controlled by $\omega R(\theta,\phi)$.
    We therefore define the inverse relative smoothing width (IRSW) as
    \begin{equation}    \label{eq:omega_Rin_cond}
        \eta \equiv \omega R_{\text{in}},
    \end{equation}
    where $R_{\text{in}}$ is the inradius, i.e.,~the minimum radial distance from the origin to the truncation boundary.
    A conservative global criterion for sTC to approximate its parent TC potential accurately is therefore $\eta \gg 1$.
    For spherical truncation, $R_{\text{in}}=R_{\text{c}}$, and the IRSW reduces to the dimensionless product $\omega R_{\text{c}}$ used above.
    For the WS truncation employed in all numerical calculations below, $R_{\text{in}}$ is the inradius of the WS cell of the BvK supercell.

    \subsection{Range separation of the sTC potential}
    \label{subsec:stc_reciprocal}

    The smoothing introduced in \cref{eq:v_stc_def} leads to a particularly simple representation of the sTC potential in reciprocal space.
    Using the convolution theorem, the Fourier transform of the sTC potential is
    \begin{equation}    \label{eq:v_stc_q}
        v_{\text{sTC}}(\bm{q}; \omega)
            = \frac{4\pi}{q^2}
            - \left[
                \frac{4\pi}{q^2}
                - v_{\text{TC}}(\bm{q})
            \right]
            \mathrm{e}^{-q^2/(4\omega^2)}.
    \end{equation}
    The Gaussian factor exponentially suppresses the large-$q$ components associated with the truncation boundary of the original TC potential.
    The full sTC potential nevertheless retains the short-range Coulomb singularity and therefore approaches $4\pi/q^2$ at large $q$.
    This behavior motivates a range separation in which the short-range Coulomb singularity is treated in real space and the smoothed TC contribution is treated in reciprocal space.

    Specifically, \cref{eq:v_stc_q} can be written as the sum of short-range (SR) and long-range (LR) components,
    \begin{equation}    \label{eq:v_stc_rs_q}
    \begin{split}
        v_{\text{sTC}}(\bm{q}; \omega)
            &= v_{\text{sTC}}^{\text{SR}}(q; \omega)
            + v_{\text{sTC}}^{\text{LR}}(\bm{q}; \omega),    \\
        v_{\text{sTC}}^{\text{SR}}(q; \omega)
            &= \frac{4\pi}{q^2}
            \left[
                1 - \mathrm{e}^{-q^2/(4\omega^2)}
            \right],    \\
        v_{\text{sTC}}^{\text{LR}}(\bm{q}; \omega)
            &= v_{\text{TC}}(\bm{q})
            \mathrm{e}^{-q^2/(4\omega^2)}.
    \end{split}
    \end{equation}
    The corresponding real-space decomposition is
    \begin{equation}    \label{eq:v_stc_rs_r}
    \begin{split}
        v_{\text{sTC}}(\bm{r}; \omega)
            &= v_{\text{sTC}}^{\text{SR}}(r; \omega)
            + v_{\text{sTC}}^{\text{LR}}(\bm{r}; \omega),    \\
        v_{\text{sTC}}^{\text{SR}}(r; \omega)
            &= \frac{\text{erfc}(\omega r)}{r},    \\
        v_{\text{sTC}}^{\text{LR}}(\bm{r}; \omega)
            &= \frac{\omega^3}{\pi^{3/2}}
            \int \mathrm{d}\bm{r}'\,
            v_{\text{TC}}(\bm{r}')
            \mathrm{e}^{-\omega^2|\bm{r}-\bm{r}'|^2}.
    \end{split}
    \end{equation}
    This decomposition is well suited for a dual-space evaluation of the corresponding Gaussian-basis ERIs, as described in \cref{subsec:eri}.

    \subsection{Electron repulsion integrals over the sTC potential}
    \label{subsec:eri}

    Evaluating the HF exchange energy with the sTC potential requires the corresponding four-center ERIs.
    In this section, we formulate a dual-space algorithm for these integrals; an efficient density-fitting implementation is presented in \cref{subsec:df}.

    We employ $n_{\text{AO}}$ periodic Gaussian-type orbitals (GTOs) at each $k$-point, constructed as Bloch sums of localized GTOs~\cite{Ye26arXiv},
    \begin{equation}    \label{eq:periodic_GTO}
        \phi_{\mu}^{\bm{k}}(\bm{r})
        = \sum_{\bm{R}} \me^{\mi \bm{k} \cdot \bm{R}} \tilde{\phi}_{\mu}^{\bm{R}}(\bm{r}),
    \end{equation}
    where $\bm{k}$ denotes a crystal momentum in the first Brillouin zone, $\bm{R}$ runs over lattice vectors, and \(
        \tilde{\phi}_{\mu}^{\bm{R}}(\bm{r})
            \equiv \tilde{\phi}_{\mu}(\bm{r}-\bm{R})
    \).
    A general four-center sTC-ERI is defined as
    \begin{equation}    \label{eq:eri_stc}
    \begin{split}
        [v_{\text{sTC}}(\omega)]_{\mu\nu\lambda\sigma}^{
        \bm{k}_1\bm{k}_2\bm{k}_3\bm{k}_4}
            &= \int_{\Omega} \mathrm{d}\bm{r}_1
            \int \mathrm{d}\bm{r}_2\,
            \rho_{\mu\nu}^{\bm{k}_1\bm{k}_2}(\bm{r}_1)
            \rho_{\lambda\sigma}^{\bm{k}_3\bm{k}_4}(\bm{r}_2) \\
            &\quad\times
            v_{\text{sTC}}(\bm{r}_{12};\omega),
    \end{split}
    \end{equation}
    where $\bm{r}_{12}=\bm{r}_2-\bm{r}_1$ and
    \begin{equation}
        \rho_{\mu\nu}^{\bm{k}_1\bm{k}_2}(\bm{r})
            = \phi_{\mu}^{\bm{k}_1*}(\bm{r})
            \phi_{\nu}^{\bm{k}_2}(\bm{r}).
    \end{equation}
    The sTC-ERI is nonzero only when crystal momentum is conserved modulo a reciprocal lattice vector,
    \begin{equation}
        -\bm{k}_1+\bm{k}_2-\bm{k}_3+\bm{k}_4
            = \bm{G}_0.
    \end{equation}

    Using the range-separated form in \cref{eq:v_stc_rs_r,eq:v_stc_rs_q}, we decompose each sTC-ERI into SR and LR contributions,
    \begin{equation}
        [v_{\text{sTC}}(\omega)]_{\mu\nu\lambda\sigma}^{
            \bm{k}_1\bm{k}_2\bm{k}_3\bm{k}_4
        }
            = [v_{\text{sTC}}^{\text{SR}}(\omega)]_{
                \mu\nu\lambda\sigma}^{
                \bm{k}_1\bm{k}_2\bm{k}_3\bm{k}_4}
            + [v_{\text{sTC}}^{\text{LR}}(\omega)]_{
                \mu\nu\lambda\sigma}^{
                \bm{k}_1\bm{k}_2\bm{k}_3\bm{k}_4}.
    \end{equation}
    Because the SR-sTC potential is exponentially localized in real space, its contribution to the ERIs can be evaluated efficiently through a real-space lattice sum,
    \begin{equation}    \label{eq:eri_stc_sr}
    \begin{split}
        [v_{\text{sTC}}^{\text{SR}}(\omega)]_{
            \mu\nu\lambda\sigma
        }^{
            \bm{k}_1\bm{k}_2\bm{k}_3\bm{k}_4
        }
            &= \sum_{\bm{R}_2\bm{R}_3\bm{R}_4}
            \me^{
                \mi (
                    \bm{k}_2\cdot\bm{R}_2
                    -\bm{k}_3\cdot\bm{R}_3
                    +\bm{k}_4\cdot\bm{R}_4
                )
            } \\
            &\quad\times
            [v^{\text{SR}}_{\text{sTC}}(\omega)]_{
                \mu\nu\lambda\sigma
            }^{\bm{0}\bm{R}_2\bm{R}_3\bm{R}_4},
    \end{split}
    \end{equation}
    where
    \begin{equation}    \label{eq:eri_stc_sr_nonpbc}
    \begin{split}
        [v^{\text{SR}}_{\text{sTC}}(\omega)]_{
            \mu\nu\lambda\sigma
        }^{\bm{0}\bm{R}_2\bm{R}_3\bm{R}_4}
            &= \iint \mathrm{d}\bm{r}_1\mathrm{d}\bm{r}_2\,
            \tilde{\phi}_{\mu}^{\bm{0}}(\bm{r}_1)
            \tilde{\phi}_{\nu}^{\bm{R}_2}(\bm{r}_1) \\
            &\quad\times
            \frac{\text{erfc}(\omega r_{12})}{r_{12}}
            \tilde{\phi}_{\lambda}^{\bm{R}_3}(\bm{r}_2)
            \tilde{\phi}_{\sigma}^{\bm{R}_4}(\bm{r}_2),
        \end{split}
    \end{equation}
    is a nonperiodic four-center ERI over the complementary-error-function-attenuated Coulomb kernel.
    Such integrals are supported by standard molecular Gaussian integral libraries~\cite{Sun15JCC,Libint2}.
    The triple lattice sum in \cref{eq:eri_stc_sr} converges rapidly and can therefore be truncated at a finite range~\cite{Sun20arXiv,Sun23arXiv}.

    The LR contribution is evaluated through a reciprocal-space lattice sum,
    \begin{equation}    \label{eq:eri_stc_lr}
    \begin{split}
        [v_{\text{sTC}}^{\text{LR}}(\omega)]_{
            \mu\nu\lambda\sigma
        }^{
            \bm{k}_1\bm{k}_2\bm{k}_3\bm{k}_4
        }
            &= \frac{1}{\Omega}
            \sum_{\bm{G}}
            \rho_{\mu\nu}^{\bm{k}_1\bm{k}_2}(\bm{G})
            \rho_{\lambda\sigma}^{\bm{k}_3\bm{k}_4}(-\bm{G}) \\
            &\quad\times
            v_{\text{TC}}(\bm{G}+\bm{k}_{12})
            \me^{-|\bm{G}+\bm{k}_{12}|^2/(4\omega^2)},
    \end{split}
    \end{equation}
    where $\bm{k}_{12}=\bm{k}_2-\bm{k}_1$ and
    \begin{equation}    \label{eq:rho_G}
        \rho_{\mu\nu}^{\bm{k}_1\bm{k}_2}(\bm{G})
            = \int_{\Omega}\mathrm{d}\bm{r}\,
            \me^{-\mi(\bm{k}_{12}+\bm{G})\cdot\bm{r}}
            \rho_{\mu\nu}^{\bm{k}_1\bm{k}_2}(\bm{r}).
    \end{equation}
    Efficient constructions of the reciprocal-space TC kernel $v_{\text{TC}}(\bm{q})$ are available from the original plane-wave formulations~\cite{Spence08PRB,Sundararaman13PRB}.
    The Gaussian damping factor in \cref{eq:eri_stc_lr} exponentially suppresses the large-$|\bm{G}|$ contributions, allowing the reciprocal-space lattice sum to be truncated at a modest plane-wave cutoff.
    This truncation remains effective for all-electron calculations: although compact core contributions to the pair densities decay slowly in reciprocal space, their contribution to the LR sTC-ERI is strongly suppressed by the Gaussian damping factor.
    To avoid explicitly resolving these compact pair densities on a fine real-space grid, we evaluate the Fourier components in \cref{eq:rho_G} analytically following ref~\onlinecite{Sun17JCP,Ye21JCPa}.

    In passing, we mention that when core electrons are replaced by a pseudopotential or effective core potential, the sTC ERIs can alternatively be evaluated entirely in reciprocal space,
    \begin{equation}    \label{eq:eri_stc_fftdf}
    \begin{split}
        [v_{\text{sTC}}(\omega)]_{
            \mu\nu\lambda\sigma
        }^{
            \bm{k}_1\bm{k}_2\bm{k}_3\bm{k}_4
        }
            &= \frac{1}{\Omega}
            \sum_{\bm{G}}
            \rho_{\mu\nu}^{\bm{k}_1\bm{k}_2}(\bm{G})
            \rho_{\lambda\sigma}^{\bm{k}_3\bm{k}_4}(-\bm{G}) \\
            &\quad\times
            v_{\text{sTC}}(\bm{G}+\bm{k}_{12})
    \end{split}
    \end{equation}
    where the reciprocal-space sTC potential is given in \cref{eq:v_stc_q}.
    This approach is termed FFTDF in PySCF and is closely related to the GPW method in CP2K.
    Because PySCF already supports FFTDF-based TC integrals, extending the implementation to sTC through \cref{eq:eri_stc_fftdf} is straightforward and provides a controlled reference for comparing the two potentials in pseudopotential calculations, as discussed in \cref{sec:results}.

    \subsection{Range-separated density fitting}
    \label{subsec:df}

    The cost of handling the four-center sTC-ERIs can be reduced through density fitting (DF)~\cite{Whitten73JCP,Dunlap00PCCP}.
    The decomposition into SR and LR components fits naturally within the range-separated density-fitting (RSDF) framework developed by one of the authors~\cite{Ye21JCPa,Ye21JCPb}.
    Using RSDF, the four-center sTC-ERIs are approximated as
    \begin{equation}    \label{eq:eri_stc_df}
        [v_{\text{sTC}}(\omega)]_{
            \mu\nu\lambda\sigma
        }^{
            \bm{k}_1\bm{k}_2\bm{k}_3\bm{k}_4
        }
            \approx \sum_{P} [B_{\text{sTC}}(\omega)]_{P\mu\nu}^{\bm{k}_1\bm{k}_2}
            [B_{\text{sTC}}(\omega)]_{P\lambda\sigma}^{\bm{k}_3\bm{k}_4},
    \end{equation}
    where $P$ labels a set of $n_{\text{aux}}$ auxiliary basis functions $\chi_{P}^{\bm{k}}(\bm{r})$, which are periodic GTOs defined as in \cref{eq:periodic_GTO}.
    The three-index DF factors are determined by solving a linear fitting equation,
    \begin{equation}    \label{eq:df_eqn}
        \sum_{P} [L_{\text{sTC}}(\omega)]_{QP}^{\bm{k}_{12}}
        [B_{\text{sTC}}(\omega)]^{\bm{k}_1\bm{k}_2}_{P\mu\nu}
            = [v_{\text{sTC}}(\omega)]^{\bm{k}_1\bm{k}_2}_{Q\mu\nu}.
    \end{equation}
    The integrals needed by \cref{eq:df_eqn} are the three-center sTC-ERIs between an auxiliary basis function and a pair density,
    \begin{equation}
        [v_{\text{sTC}}(\omega)]^{\bm{k}_1\bm{k}_2}_{P\mu\nu}
            = \int_{\Omega} \mathrm{d}\bm{r}_1
            \int \mathrm{d}\bm{r}_2\,
            \chi_{P}^{\bm{k}_{12}*}(\bm{r}_1)
            \rho_{\mu\nu}^{\bm{k}_1\bm{k}_2}(\bm{r}_2)
            v_{\text{sTC}}(\bm{r}_{12}; \omega),
    \end{equation}
    and the two-center sTC-ERIs between two auxiliary basis functions,
    \begin{equation}    \label{eq:eri_stc_2c}
        [v_{\text{sTC}}(\omega)]^{\bm{k}}_{PQ}
            = \int_{\Omega} \mathrm{d}\bm{r}_1
            \int \mathrm{d}\bm{r}_2\,
            \chi_{P}^{\bm{k}*}(\bm{r}_1)
            \chi_{Q}^{\bm{k}}(\bm{r}_2)
            v_{\text{sTC}}(\bm{r}_{12}; \omega),
    \end{equation}
    where $[\mathbf{L}_{\text{sTC}}(\omega)]^{\bm{k}}$ is the lower triangular Cholesky factor of the two-center sTC-ERI matrix at crystal momentum $\bm{k}$,
    \begin{equation}    \label{eq:eri_stc_metric_cholesky}
        [\mathbf{v}_{\text{sTC}}(\omega)]^{\bm{k}}
            = [\mathbf{L}_{\text{sTC}}(\omega)]^{\bm{k}}
            [\mathbf{L}_{\text{sTC}}(\omega)]^{\bm{k}\dagger}.
    \end{equation}
    Following ref~\onlinecite{Ye21JCPa}, both the two-center and three-center sTC-ERIs can be evaluated using the dual-space algorithm outlined in \cref{subsec:eri}.
    This RSDF formulation does not assume a particular treatment of the nuclei and is therefore applicable to both pseudopotential and all-electron calculations~\cite{Ye21JCPa}.
    RSDF reduces the storage scaling from $\mathcal{O}(N_k^3 n_{\text{AO}}^4)$ for four-center sTC-ERIs to $\mathcal{O}(N_k^2 n_{\text{AO}}^2 n_{\text{aux}})$ for three-center sTC-ERIs and DF factors, which is sufficient for the pilot benchmarks presented in this work.

    \subsection{Choice of the smoothing parameter \texorpdfstring{$\omega$}{omega}}
    \label{subsec:omega}

    As discussed in \cref{subsec:stc}, the smoothing parameter $\omega$ is most conveniently specified through the IRSW $\eta=\omega R_{\text{in}}$, which must be sufficiently large for sTC to approximate its parent TC potential accurately.
    In RSDF, $\omega$ also controls the balance between the SR real-space and LR reciprocal-space ERI evaluations and should therefore remain within the practical range $\omega \simeq 0.1$--$1~\text{Bohr}^{-1}$ suggested by ref~\onlinecite{Ye21JCPa}.
    For a sequence of uniformly enlarged, fixed-shape $k$-point meshes, the truncation boundary scales with the BvK supercell.
    At fixed $\eta$,
    \begin{equation}    \label{eq:omega_scaling}
        \omega
            \sim R_{\text{in}}^{-1}
            \sim N_k^{-1/3}.
    \end{equation}
    Thus, $\omega$ decreases as the TDL is approached, lowering the plane-wave cutoff required for the LR component while increasing the spatial extent of the SR component.
    A practical question is therefore whether values of $\eta$ that reproduce TC accurately also keep $\omega$ within the established RSDF working range for the $k$-meshes used in realistic calculations.
    This question is examined numerically in \cref{sec:results}.

    \section{Computational details}
    \label{sec:comp_details}

    We implemented sTC-based periodic HF calculations using both FFTDF and RSDF in a development version of PySCF~\cite{Sun18WIRCMS,Sun20JCP,Sun26JCP}.
    The FFTDF implementation closely follows the existing TC implementation in PySCF and enables direct sTC--TC comparisons in pseudopotential calculations, which we use to benchmark convergence with respect to $\eta$.
    The RSDF implementation extends these tests to all-electron calculations, for which direct TC reference calculations are computationally impractical.
    Although our implementation supports both spherical and WS truncation boundaries, we report only WS-truncation results because the nonspherical boundary provides a more stringent test of the smoothing approximation.

    Our benchmarks span three classes of materials with diverse electronic structures: three gapped solids [diamond, silicon, and hexagonal boron nitride (h-BN)], two metallic solids (lithium and aluminum), and three molecular crystals [carbon dioxide, ammonia, and tetrathiafulvalene (TTF)].
    The lattice structures are reported in the Supporting Information.

    For pseudopotential calculations, we use Goedecker--Teter--Hutter (GTH) pseudopotentials~\cite{Goedecker96PRB,Hartwigsen98PRB,HutterPP} with GTH-cc-pVDZ basis sets~\cite{Ye22JCTC}, except for h-BN, for which a minimal basis set enables calculations on larger $k$-meshes.
    For the diamond band structure calculations, the GTH-cc-pVTZ basis set is used.
    The plane-wave basis used in FFTDF is determined from kinetic-energy cutoffs of $100~E_{\text{h}}$ for diamond, $40~E_{\text{h}}$ for silicon, $100~E_{\text{h}}$ for h-BN, $125~E_{\text{h}}$ for lithium, and $40~E_{\text{h}}$ for aluminum, which converge the final HF exchange energy to ca.~$10^{-8}~E_{\text{h}}$ per atom.
    For all-electron calculations, we use Dunning's cc-pVDZ basis sets~\cite{Dunning89JCP} for diamond and silicon and the MINAO basis set for h-BN and the three molecular crystals.
    The cc-pVDZ-JKFIT auxiliary basis sets~\cite{Weigend02PCCP} are used for the RSDF evaluation of ERIs for diamond and silicon, whereas an even-tempered auxiliary basis with a progression factor of $\beta = 2$ is used for MINAO-based calculations.
    For lithium and aluminum, the standard cc-pVDZ basis sets contain diffuse basis functions that cause severe linear-dependence problems over the $k$-meshes studied here.
    We generated numerically stable cc-pVDZ basis sets and the corresponding auxiliary basis sets for RSDF using the material-constrained atomic optimization algorithm (MCAO) from ref~\onlinecite{Yu26arXiv}.

    For most calculations, we use isotropic $n \times n \times n$ $k$-point meshes to sample the first Brillouin zone, with $N_k = n^3$ denoting the total number of sampled $k$-points.
    \myred{For h-BN and TTF, which have strongly anisotropic unit cells, we also perform calculations using anisotropic $k$-meshes---$3n \times 3n \times n$ for h-BN and $n \times 2n \times n$ for TTF---that compensate for the unit-cell anisotropy and make the corresponding BvK supercells nearly isotropic.}
    The $k$-meshes for all gapped materials are unshifted and include the $\Gamma$-point.
    For lithium and aluminum, the $k$-meshes are shifted by the Baldereschi mean-value points~\cite{Baldereschi73PRB,Drummond26arXiv} $(\frac{1}{6},\frac{1}{6},\frac{1}{2})$ and $(0.1477, 0.3112, 0.4588)$, respectively, expressed in reciprocal lattice coordinates of the corresponding BvK supercell.

    \myred{To evaluate the FSE on finite $k$-meshes, we obtain the TDL reference HF exchange energy $E_{\text{HFX}}^{\text{TDL}}$ for each system by extrapolating the probe-charge Ewald results.
    For gapped systems, we fit results from the three densest $k$-meshes to the following quadratic form~\cite{Goldzak22JCP} based on the $\mathcal{O}(N_k^{-1})$ leading-order FSE scaling~\cite{Broqvist09PRB,Xing24PRX},
    \begin{equation}    \label{eq:tdl_gapped}
        E_{\text{HFX}}(N_k)
            = E_{\text{HFX}}^{\text{TDL}} + A N_k^{-1} + B N_k^{-2}.
    \end{equation}
    For the molecular crystals, the same extrapolation formula is applied to the HF total energy per molecule.
    For the two strongly anisotropic systems, h-BN and TTF, we use the anisotropic $k$-meshes described above.
    For metallic systems, we perform a similar three-point extrapolation based on the $\mathcal{O}(N_k^{-2/3})$ leading-order FSE scaling~\cite{Neufeld23PRL},
    \begin{equation}    \label{eq:tdl_metal}
        E_{\text{HFX}}(N_k)
            = E_{\text{HFX}}^{\text{TDL}} + A N_k^{-2/3} + B N_k^{-1}.
    \end{equation}
    As discussed in \cref{subsec:tdl_extrapolation} and the Supporting Information, we estimate the accuracy of these TDL extrapolations to be \textit{ca}.~$0.3$~$\text{m}E_{\text{h}}$ per atom for the bulk semiconductors and metals and \textit{ca}.~$0.1$~kcal/mol per molecule for the molecular crystals.
    This accuracy is sufficient for the purposes of this work.
    }
    The TDL reference lattice constant and bulk modulus of diamond are obtained by first extrapolating the probe-charge Ewald HF energy at each volume and then fitting the extrapolated energy--volume curve to the Birch--Murnaghan equation of state.
    The diamond band structures were calculated using a $4\times4\times4$ $k$-mesh without extrapolation.

    For all gapped systems, including the three molecular crystals, the results for each FSE treatment are obtained from independent self-consistent-field (SCF) calculations.
    For the metallic systems, different FSE treatments can converge to distinct but energetically nearly degenerate SCF solutions especially as the TDL is approached, complicating direct comparisons of their exchange energies.
    To isolate the effect of the FSE treatment, we follow previous work~\cite{Spence08PRB,Sundararaman13PRB} and evaluate $E_{\text{HFX}}$ non-self-consistently for each case using a common HF density matrix obtained from the probe-charge Ewald SCF calculation at the same $k$-point mesh.

    \section{Results and Discussion}
    \label{sec:results}

    The numerical tests address three questions: how rapidly WS-sTC approaches WS-TC as $\eta$ is increased; whether the corresponding values of $\omega=\eta/R_{\text{in}}$ remain practical for all-electron RSDF calculations; and whether sTC preserves TC-quality energies, structural properties, and band structures across gapped, metallic, and molecular-crystal systems.
    We consider sTC accurate at a given $k$-mesh when the direct sTC--TC difference is negligible relative to the remaining TC finite-size error.

    \subsection{Gapped systems}
    \label{subsec:gapped_systems}

    \begin{figure}[!t]
        \centering
        \includegraphics[width=3.4in]{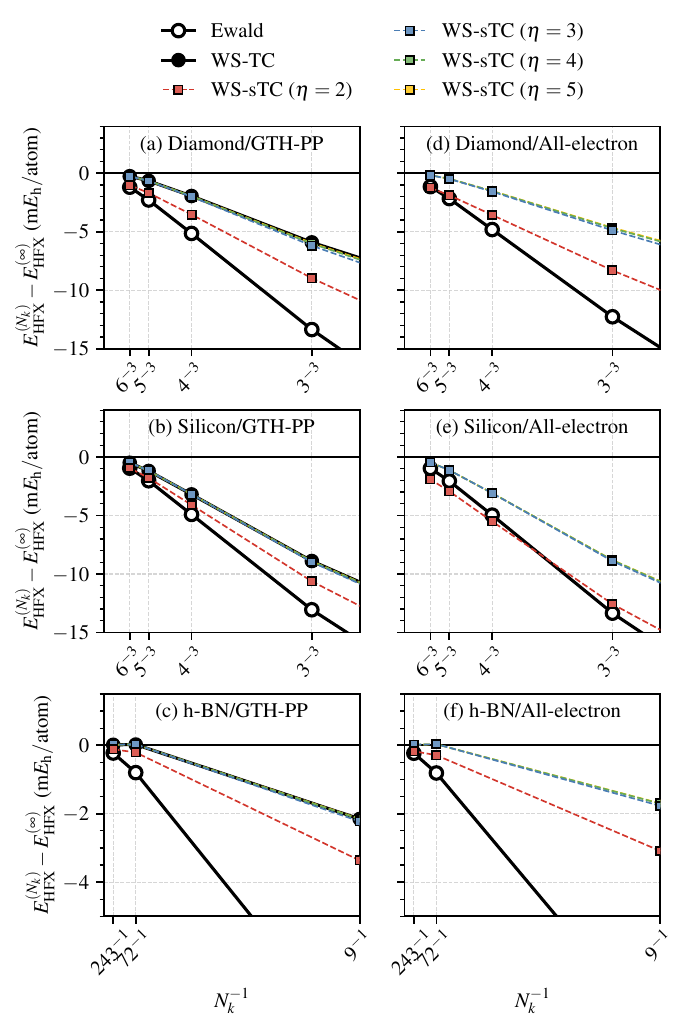}
        \caption{Finite-size error in the HF exchange energy ($E_{\text{HFX}}$) for three gapped materials obtained using WS-sTC with different values of $\eta$, WS-TC, and the probe-charge Ewald method.
        The left column (a--c) shows results obtained using GTH pseudopotentials with FFTDF, while the right column (d--f) shows the corresponding all-electron results obtained using RSDF.
        The WS-TC results are included for pseudopotential-based calculations only.
        }
        \label{fig:C_Si_hBN_ek_conv_Nk}
    \end{figure}

    \Cref{fig:C_Si_hBN_ek_conv_Nk}(a--c) compares the FSE in pseudopotential-based $E_{\text{HFX}}$ for the three gapped systems obtained using sTC with $\eta = 2$--$5$, TC, and probe-charge Ewald.
    As discussed in \cref{subsec:stc}, increasing $\eta$ narrows the smoothing region around the truncation boundary and drives sTC toward its parent TC potential.
    For the loose choice of $\eta = 2$, the sTC curve remains noticeably separated from the TC reference, although it already follows the same qualitative FSE decay.
    Increasing $\eta$ to $3$ brings the sTC and TC FSE curves into close agreement, and the agreement improves further at $\eta=4$ and $5$.
    As quantified directly in Fig.~S1 of the Supporting Information, the sTC--TC difference for $\eta \geq 3$ on $4\times4\times4$ or larger $k$-meshes is well below $0.1~\text{m}E_{\text{h}}$ per atom, approximately two orders of magnitude smaller than the remaining TC FSE.
    Consequently, sTC with $\eta \geq 3$ preserves the rapid TDL convergence of TC and yields a smaller FSE than the probe-charge Ewald method at each $k$-mesh examined.
    This is observed for both isotropic diamond and silicon and anisotropic h-BN, which supports our inradius-based definition of $\eta$ in \cref{eq:omega_Rin_cond} for the WS domains examined here.
    \myred{The comparatively irregular TDL convergence of h-BN obtained with isotropic $k$-meshes (Fig.~S2 of the Supporting Information) is reduced by using anisotropic meshes that make the corresponding BvK supercells nearly isotropic, as shown in \cref{fig:C_Si_hBN_ek_conv_Nk}(c,f).}
    Both mesh sequences nevertheless exhibit the same rapid convergence with respect to $\eta$.

    Having established the accuracy of sTC for the pseudopotential calculations, we next consider the main practical target of our dual-space implementation: all-electron WS-sTC calculations via RSDF.
    As shown in Fig.~S3, $\eta = 3$--$5$ corresponds to $\omega \simeq 0.1$--$0.7~\text{Bohr}^{-1}$ for the three gapped materials over the $k$-meshes examined, well within the recommended RSDF range of $\omega \simeq 0.1$--$1~\text{Bohr}^{-1}$ reported in ref~\onlinecite{Ye21JCPa}.
    Thus, the values of $\eta$ needed to reproduce pseudopotential-based TC-HF calculations do not force $\omega$ outside the standard RSDF working regime for these systems.
    The remaining question is whether, in the absence of a practical all-electron WS-TC reference, the all-electron sTC calculations retain the rapid $\eta$-convergence and advantageous FSE scaling observed in their pseudopotential counterparts.

    \Cref{fig:C_Si_hBN_ek_conv_Nk}(d--f) presents the all-electron $E_{\text{HFX}}$ FSE obtained using WS-sTC with $\eta = 2$--$5$ for the three gapped systems.
    The dependence on both $\eta$ and $k$-mesh size closely parallels that of the corresponding pseudopotential results in panels (a--c).
    In particular, $\eta = 3$ essentially converges the all-electron sTC FSE curves, which approach the TDL more rapidly than the probe-charge Ewald results.
    The agreement between the pseudopotential and all-electron trends supports the transferability of $\eta$ between the two treatments and demonstrates the practical value of the all-electron sTC implementation.

    \begin{figure}[!b]
        \centering
        \includegraphics[width=3.4in]{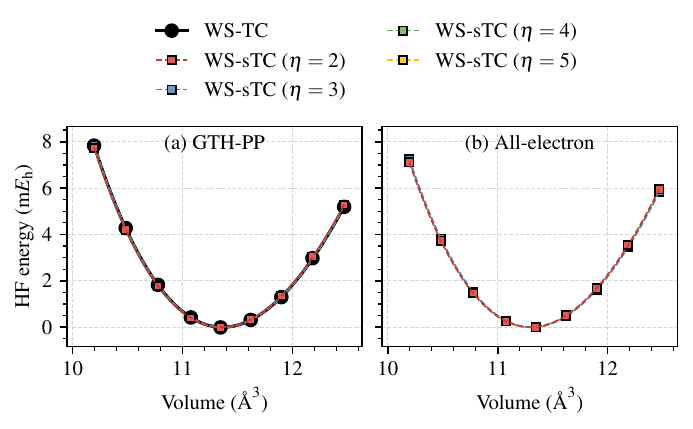}
        \caption{HF equation-of-state curves of diamond on a $5\times5\times5$ $k$-point mesh.
        Panel (a) compares WS-sTC with $\eta=2$--$5$ against WS-TC using the GTH pseudopotential and GTH-cc-pVDZ basis set.
        Panel (b) shows all-electron WS-sTC results with $\eta=2$--$5$ using the cc-pVDZ basis set; a direct all-electron WS-TC reference is not included because it is computationally impractical.
        For each method, markers denote the data and lines denote the fitted Birch--Murnaghan equation of state.
        The HF energy is shifted by the fitted energy minimum separately for each method.}
        \label{fig:C_eos}
    \end{figure}

    The exchange-energy FSE benchmarks above assess sTC at fixed lattice geometries.
    We next test whether the accuracy of sTC is preserved for relative energies along a potential energy surface.
    \Cref{fig:C_eos}(a) shows the GTH-pseudopotential HF energy--volume curves of diamond over a $\pm10\%$ range around the equilibrium volume using a $5\times5\times5$ $k$-point mesh.
    Even at $\eta = 2$, the sTC curve closely reproduces the TC reference, indicating that relative energies converge more rapidly with $\eta$ than absolute energies.
    Using a tighter $\eta \geq 3$ renders the sTC and TC curves visually indistinguishable.
    In the absence of a direct all-electron TC reference, \cref{fig:C_eos}(b) shows that the all-electron sTC curves closely parallel their pseudopotential counterparts and are essentially converged with $\eta = 3$.

    \begin{table*}[!t]
        \centering
        \caption{Equilibrium lattice constant and bulk modulus of diamond calculated using WS-sTC with different values of $\eta$.
        The pseudopotential results obtained using the GTH pseudopotential and GTH-cc-pVDZ basis set are compared with WS-TC and the probe-charge Ewald method, while the all-electron results obtained using the cc-pVDZ basis set are compared with the probe-charge Ewald method.
        \myred{The TDL results are obtained by first extrapolating the HF energy at each volume using \cref{eq:tdl_gapped} for probe-charge Ewald and \cref{eq:tdl_exp} for WS-TC and WS-sTC} and then performing the Birch--Murnaghan fit.}
        \label{tab:C_a0_B0_conv}
        \begin{tabular}{llccccccccccccc}
            \hline\hline
            & & \multirow{2}*{$k$-mesh} &
                & \multicolumn{5}{c}{Lattice constant ($\text{\AA}$)} &
                & \multicolumn{5}{c}{Bulk modulus (GPa)} \\
                \cmidrule(lr){5-9}
                \cmidrule(lr){11-15}
            & &   &
                & sTC ($\eta=2$) & sTC ($\eta=3$) & sTC ($\eta=4$) & TC & Ewald &
                & sTC ($\eta=2$) & sTC ($\eta=3$) & sTC ($\eta=4$) & TC & Ewald \\
            \hline
            \multirow{5}{*}{\rotatebox[origin=c]{90}{GTH-PP}} & $\phantom{11}$ &
            $3\times3\times3$ & $\phantom{11}$ & $3.580$ & $3.582$ & $3.582$ & $3.583$ & $3.575$ & $\phantom{11}$ & $478.9$ & $477.5$ & $477.4$ & $477.4$ & $481.6$  \\
             & $\phantom{11}$ & $4\times4\times4$ & $\phantom{11}$ & $3.570$ & $3.571$ & $3.571$ & $3.571$ & $3.568$ & $\phantom{11}$ & $483.4$ & $482.6$ & $482.6$ & $482.6$ & $484.5$  \\
             & $\phantom{11}$ & $5\times5\times5$ & $\phantom{11}$ & $3.568$ & $3.569$ & $3.569$ & $3.569$ & $3.567$ & $\phantom{11}$ & $484.8$ & $484.2$ & $484.2$ & $484.2$ & $485.1$  \\
             & $\phantom{11}$ & $6\times6\times6$ & $\phantom{11}$ & $3.568$ & $3.568$ & $3.568$ & $3.568$ & $3.567$ & $\phantom{11}$ & $484.7$ & $484.3$ & $484.3$ & $484.1$ & $484.9$  \\
             & $\phantom{11}$ & TDL extrap. & $\phantom{11}$ & $3.568$ & $3.568$ & $3.568$ & $3.568$ & $3.568$ & $\phantom{11}$ & $484.6$ & $484.2$ & $484.2$ & $483.9$ & $484.0$  \\
            \hline
            \hline
            \multirow{5}{*}{\rotatebox[origin=c]{90}{All-electron}} & $\phantom{11}$ &
            $3\times3\times3$ & $\phantom{11}$ & $3.572$ & $3.575$ & $3.575$ & \textemdash & $3.568$ & $\phantom{11}$ & $490.8$ & $488.9$ & $488.8$ & \textemdash & $493.4$  \\
             & $\phantom{11}$ & $4\times4\times4$ & $\phantom{11}$ & $3.564$ & $3.565$ & $3.565$ & \textemdash & $3.562$ & $\phantom{11}$ & $494.6$ & $493.4$ & $493.4$ & \textemdash & $495.5$  \\
             & $\phantom{11}$ & $5\times5\times5$ & $\phantom{11}$ & $3.562$ & $3.563$ & $3.563$ & \textemdash & $3.561$ & $\phantom{11}$ & $495.1$ & $494.3$ & $494.3$ & \textemdash & $495.4$  \\
             & $\phantom{11}$ & $6\times6\times6$ & $\phantom{11}$ & $3.562$ & $3.562$ & $3.562$ & \textemdash & $3.562$ & $\phantom{11}$ & $495.1$ & $494.5$ & $494.5$ & \textemdash & $495.1$  \\
             & $\phantom{11}$ & TDL extrap. & $\phantom{11}$ & $3.562$ & $3.562$ & $3.562$ & \textemdash & $3.562$ & $\phantom{11}$ & $495.1$ & $494.5$ & $494.5$ & \textemdash & $494.4$  \\
            \hline
        \end{tabular}
    \end{table*}

    \Cref{tab:C_a0_B0_conv} reports the equilibrium lattice constants $a_0$ and bulk moduli $B_0$ obtained from Birch--Murnaghan fits.
    In the pseudopotential case, the loosest choice, $\eta = 2$, reproduces the TC reference values within $0.003~\text{\AA}$ for $a_0$ and $1.5$~GPa for $B_0$ across all $k$-meshes examined, whereas $\eta \geq 3$ reduces these differences to at most $0.001~\text{\AA}$ and $0.2$~GPa.
    For the all-electron calculations, the values obtained with $\eta=3$ and $4$ agree within $0.001~\text{\AA}$ and $0.1$~GPa across all $k$-meshes, providing a corresponding internal convergence test in the absence of all-electron TC reference values.
    At the largest $k$-mesh, the $\eta$-converged sTC results agree with the extrapolated probe-charge Ewald TDL values within $0.001~\text{\AA}$ and $0.2$~GPa for both pseudopotential and all-electron calculations.
    These results show that the rapid $\eta$-convergence of the energy--volume curves carries over to the derived structural properties and further supports the transferability of $\eta$ from pseudopotential to all-electron calculations.

    \begin{figure}[!t]
        \centering
        \includegraphics[width=3.4in]{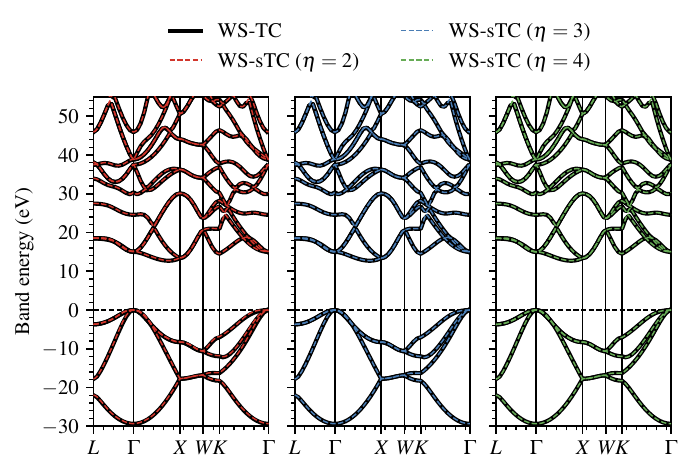}
        \caption{HF band structure of diamond calculated using WS-sTC with $\eta=2$--$4$ and WS-TC on a $4\times4\times4$ $k$-point mesh.
        The GTH pseudopotential and GTH-cc-pVTZ basis set are used.
        For each calculation, the band energies are shifted relative to its valence-band maximum.
        }
        \label{fig:C_bands}
    \end{figure}

    \Cref{fig:C_bands} further compares the sTC and TC GTH-pseudopotential HF band structures of diamond over an energy window from $-30$ to $55$~eV relative to the respective valence-band maxima.
    The corresponding band structures for silicon are shown in Fig.~S4 in the Supporting Information.
    For both systems, the TC-HF bands are already reproduced closely by sTC at the loosest $\eta=2$, and the agreement improves further as $\eta$ is increased.
    As quantified in \cref{tab:gap}, the direct sTC-HF gap for diamond at the $\Gamma$ point differs from the TC reference by $0.13$, $0.05$, and $0.02$~eV for $\eta=2$, $3$, and $4$, respectively.
    The corresponding indirect gaps agree with TC to the reported precision for all three values of $\eta$.
    The close agreement across both occupied and virtual bands shows that the smoothing has little effect on the HF spectral properties at the recommended $\eta$ values.

    \begin{table}[!t]
        \centering
        \caption{Direct (at $\Gamma$) and indirect HF band gaps (in eV) of diamond calculated using WS-sTC with different values of $\eta$ and WS-TC on a $4 \times 4 \times 4$ $k$-point mesh.}
        \label{tab:gap}
        \begin{tabular}{lcccc}
            \hline\hline
            & sTC ($\eta=2$) & sTC ($\eta=3$) & sTC ($\eta=4$) & TC \\
            \hline
            Direct gap & $15.13$ & $15.05$ & $15.02$ & $15.00$ \\
            Indirect gap & $12.73$ & $12.73$ & $12.73$ & $12.73$ \\
            \hline
        \end{tabular}
    \end{table}

    \subsection{Metallic systems}
    \label{subsec:metals}

    The rapid convergence of sTC with respect to $\eta$ for gapped systems can be understood from the exponential localization of their Wannier functions~\cite{He01PRL}, which reduces the sensitivity of HF exchange to the truncation boundary.
    In metals, the one-particle density matrix instead exhibits only algebraic decay at zero temperature~\cite{IsmailBeigi99PRL,Taraskin02PRB}, and a slower convergence of sTC to the TC reference is therefore expected.
    We examine whether the values of $\eta$ identified for gapped systems nevertheless remain practically useful for metallic systems.

    \begin{figure}[!t]
        \centering
        \includegraphics[width=3.4in]{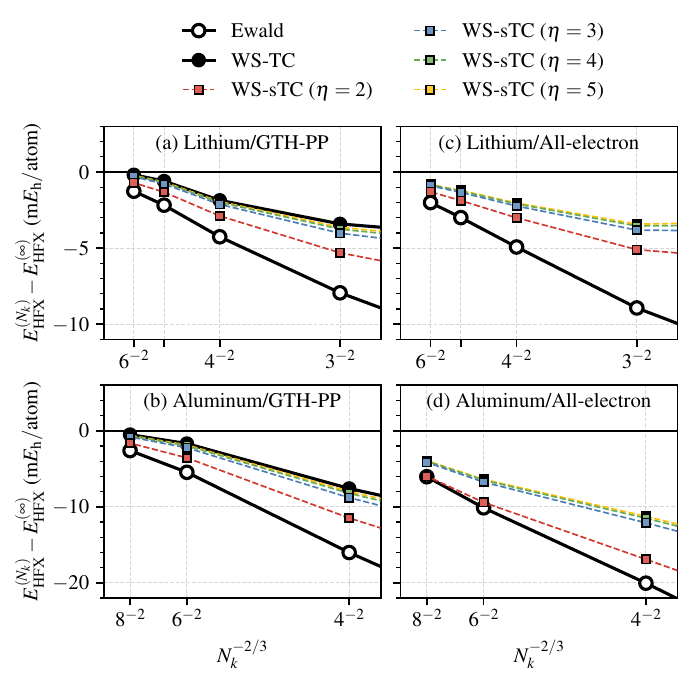}
        \caption{Finite-size error in the HF exchange energy ($E_{\text{HFX}}$) for two metallic solids obtained using WS-sTC with different values of $\eta$, WS-TC, and the probe-charge Ewald method.
        The left column (a,b) shows results obtained using GTH pseudopotentials with FFTDF, while the right column (c,d) shows the corresponding all-electron results obtained using RSDF.
        The WS-TC results are included for pseudopotential-based calculations only.}
        \label{fig:Li_Al_ek_conv_Nk}
    \end{figure}

    \Cref{fig:Li_Al_ek_conv_Nk}(a,b) compares the FSE in pseudopotential-based $E_{\text{HFX}}$ obtained using sTC with $\eta = 2$--$5$, TC, and the probe-charge Ewald method for the two metallic systems.
    As $\eta$ increases, the sTC FSE curves approach the TC reference curve monotonically for both metals, but the convergence rate is noticeably slower than that observed in \cref{fig:C_Si_hBN_ek_conv_Nk} for gapped systems.
    As quantified in Fig.~S5, the direct sTC--TC difference for $\eta \geq 3$ is $0.1$--$1~\text{m}E_{\text{h}}$ per atom for both metals, approximately one order of magnitude larger than the corresponding errors for the gapped systems.
    Nevertheless, for $\eta \geq 4$, the sTC FSE curves closely follow TC and improve upon the probe-charge Ewald results at every $k$-mesh examined.
    \Cref{fig:Li_Al_ek_conv_Nk}(c,d) further demonstrates that all-electron sTC calculations exhibit the same convergence trends with respect to $\eta$.
    Even at $\eta=3$, all-electron sTC exchange energies approach the TDL more rapidly than the probe-charge Ewald results, whereas $\eta=4$ more closely reproduces the pseudopotential TC benchmark.
    For lithium and aluminum, $\eta=3$--$5$ yields $\omega \simeq 0.1$--$0.5~\text{Bohr}^{-1}$ over the $k$-meshes examined (Fig.~S3), well within the established RSDF working range.

    \subsection{Molecular crystals}
    \label{subsec:molcrys}

    Accurate lattice energies of molecular solids are important in pharmaceutical and energy-related applications~\cite{Ostroverkhova16CR,Ma23JPS}.
    Reaching chemical accuracy in lattice energy prediction (an error below $1$~kcal/mol) typically requires hybrid DFT~\cite{Hoja19SA} or higher-level methods~\cite{Liang23JPCL,DellaPia24PRL,Syty25JCTC,Shi26arXiv}, for which the slow TDL convergence of HF exchange can become a computational bottleneck.
    We therefore examine whether sTC offers a practical advantage over the probe-charge Ewald method for HF lattice energies.

    \begin{figure}[!h]
        \centering
        \includegraphics[width=3.4in]{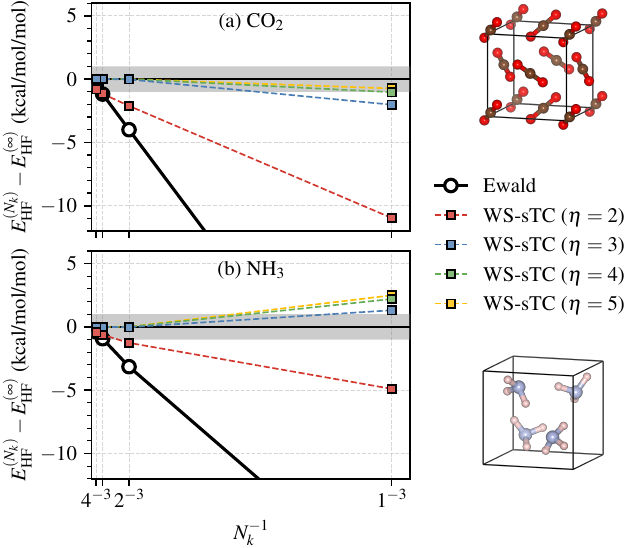}
        \caption{Finite-size error in the per-molecule HF total energy ($E_{\text{HF}}$) of (a) \ce{CO2} and (b) \ce{NH3} molecular crystals obtained using WS-sTC with different values of $\eta$ and the probe-charge Ewald method.
        All calculations were performed using all-electron RSDF.
        The gray shaded area indicates FSEs within $\pm 1$~kcal/mol.
        The unit-cell structures of the two crystals are visualized on the right (C: brown; O: red; N: blue; H: pink).
        }
        \label{fig:CO2_NH3_ek_conv_Nk}
    \end{figure}

    The lattice energy of a molecular crystal is defined as the energy difference between a molecule in the crystalline environment and in the gas phase.
    Because the gas-phase contribution is obtained from a nonperiodic molecular calculation, the dependence on the periodic FSE treatment is contained entirely in the per-molecule crystal energy.
    We therefore use the FSE in this quantity as a proxy for the finite-size contribution to the lattice-energy error.
    \Cref{fig:CO2_NH3_ek_conv_Nk} reports this error as a function of $k$-mesh size for the \ce{CO2} and \ce{NH3} crystals, calculated using all-electron RSDF with sTC ($\eta = 2$--$5$) and the probe-charge Ewald method.
    Both crystals have nearly isotropic unit cells but different dominant intermolecular interactions: \ce{CO2} is nonpolar and dominated by quadrupole--quadrupole interactions, whereas \ce{NH3} is polar and dominated by dipole--dipole interactions.
    As in the gapped systems studied above, $\eta = 2$ gives unconverged sTC results with a noticeable smoothing error, whereas $\eta = 3$ yields FSE curves that are indistinguishable from those obtained with larger $\eta$ on $2\times2\times2$ or denser $k$-meshes.
    The $\eta$-converged sTC results approach the TDL substantially faster than the probe-charge Ewald approach: $\Gamma$-point sampling already yields errors near chemical accuracy for both crystals, and a $2\times2\times2$ $k$-mesh agrees closely with the extrapolated TDL reference without further extrapolation.
    By contrast, the probe-charge Ewald results do not reach chemical accuracy until a $3\times3\times3$ $k$-mesh is used and therefore require extrapolation for accurate lattice-energy predictions on smaller meshes~\cite{Bintrim22JCTC,Liang23JPCL}.

    \begin{figure}[!h]
        \centering
        \includegraphics[width=3.4in]{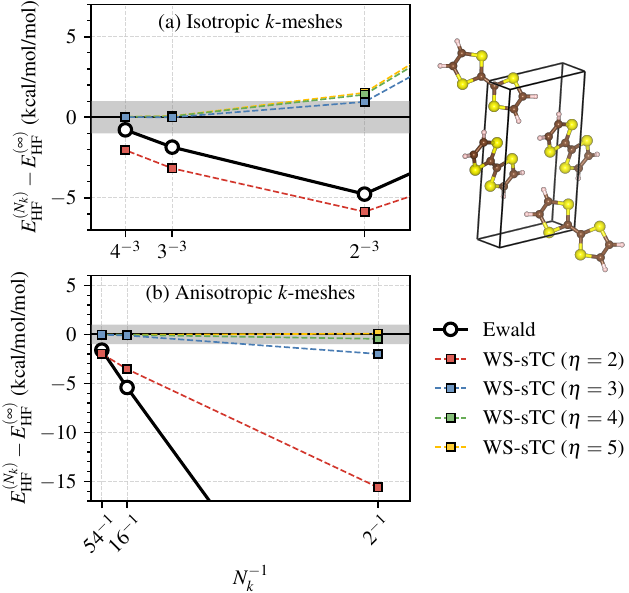}
        \caption{Finite-size error in the per-molecule HF total energy ($E_{\text{HF}}$) of the TTF crystal obtained using WS-sTC with different values of $\eta$ and the probe-charge Ewald method.
        Panel (a) uses isotropic $n \times n \times n$ $k$-point meshes, which preserve the anisotropy of the unit cell in the corresponding BvK supercells, whereas panel (b) uses anisotropic $n \times 2n \times n$ meshes, which make the BvK supercells more isotropic.
        All calculations were performed using all-electron RSDF.
        The gray shaded area indicates FSEs within $\pm 1$~kcal/mol.
        The unit-cell structure of the TTF crystal is visualized on the right (C: brown; S: yellow; H: pink).
        }
        \label{fig:TTF_TTF_ek_conv_Nk}
    \end{figure}

    Finally, we use the TTF molecular crystal as a stringent test of cell-shape effects on sTC.
    The TTF crystal has a large, anisotropic unit cell with a substantially shorter $b$ lattice vector than $a$ and $c$.
    We therefore compare an isotropic $n \times n \times n$ $k$-mesh sequence, which preserves this anisotropy in the corresponding BvK supercells, with an anisotropic $n \times 2n \times n$ sequence that makes the BvK supercells more isotropic.
    For both sequences, the HF FSE convergence with respect to $\eta$ in \cref{fig:TTF_TTF_ek_conv_Nk} follows the same qualitative trend observed for \ce{CO2} and \ce{NH3}: $\eta=2$ produces a sizable smoothing error, $\eta=3$ removes most of this error, and $\eta=4$ converges the sTC FSE to within $1$~kcal/mol of the $\eta=5$ result.
    Matching the BvK-supercell shape to the unit-cell anisotropy markedly accelerates convergence toward the TDL.
    In particular, the $\eta$-converged sTC result obtained on the sparse $1 \times 2 \times 1$ mesh already lies within chemical accuracy, whereas the corresponding isotropic sequence requires denser sampling.
    For both mesh sequences, sTC with $\eta \geq 3$ also yields substantially smaller FSEs than the probe-charge Ewald method, demonstrating that the advantage of sTC is retained for large, anisotropic molecular crystals.

    \subsection{Extrapolation to the thermodynamic limit}
    \label{subsec:tdl_extrapolation}

    \begin{table}[!h]
        \centering
        \caption{\myred{TDL-extrapolated per-atom HF exchange energy ($E_{\text{HFX}}^{\text{TDL}}$) and the associated FSE for diamond obtained using probe-charge Ewald, WS-TC, and WS-sTC with $\eta = 3$--$5$.
        Results are shown for both GTH-PP calculations with the GTH-cc-pVDZ basis set and all-electron calculations with the cc-pVDZ basis set.
        For GTH-PP calculations, the WS-TC $E_{\text{HFX}}^{\text{TDL}}$ obtained from the $(4^3,5^3,6^3)$ extrapolation is used as the reference for calculating the FSE.
        For all-electron calculations, the corresponding WS-sTC result with $\eta = 5$ is used as the reference.
        The parenthesized lists give $N_k$ for each mesh used in the extrapolation.
        The FSEs are signed differences from these references, whose entries are marked ``TDL ref.''; entries not calculated are denoted by an em dash.}
        }
        \label{tab:C_tdl}
        \begin{tabular}{llcccc}
            \hline\hline
            \multirow{2}*{Method} & \multirow{2}*{$k$-meshes}
                & \multicolumn{2}{c}{GTH-PP}
                & \multicolumn{2}{c}{All-electron}  \\
            \cmidrule(lr){3-4} \cmidrule(lr){5-6}
                &
                & $E_{\text{HFX}}^{\text{TDL}}/E_{\text{h}}$
                & FSE$/\text{m}E_{\text{h}}$
                & $E_{\text{HFX}}^{\text{TDL}}/E_{\text{h}}$
                & FSE$/\text{m}E_{\text{h}}$    \\
            \hline
            \multirow{3}{4em}{Ewald} & ($3^3, 4^3, 5^3$) & $-1.61274$ & $0.75$ & $-5.24284$ & $0.64$ \\
             & ($4^3, 5^3, 6^3$) & $-1.61328$ & $0.21$ & $-5.24332$ & $0.15$ \\
             & ($5^3, 6^3, 7^3$) & $-1.61341$ & $0.07$ & $-5.24344$ & $0.04$ \\
            \hline
            \multirow{2}{4em}{WS-TC} & ($3^3, 4^3, 5^3$) & $-1.61344$ & $0.05$ & \textemdash & \textemdash \\
             & ($4^3, 5^3, 6^3$) & $-1.61349$ & TDL ref. & \textemdash & \textemdash \\
            \hline
            \multirow{2}{4em}{WS-sTC ($\eta=3$)} & ($3^3, 4^3, 5^3$) & $-1.61345$ & $0.03$ & $-5.24341$ & $0.07$ \\
             & ($4^3, 5^3, 6^3$) & $-1.61349$ & $0.00$ & $-5.24348$ & $0.00$ \\
            \hline
            \multirow{2}{4em}{WS-sTC ($\eta=4$)} & ($3^3, 4^3, 5^3$) & $-1.61345$ & $0.04$ & $-5.24340$ & $0.08$ \\
             & ($4^3, 5^3, 6^3$) & $-1.61349$ & $0.00$ & $-5.24348$ & $0.00$ \\
            \hline
            \multirow{2}{4em}{WS-sTC ($\eta=5$)} & ($3^3, 4^3, 5^3$) & $-1.61345$ & $0.04$ & $-5.24340$ & $0.08$ \\
             & ($4^3, 5^3, 6^3$) & $-1.61349$ & $0.00$ & $-5.24348$ & TDL ref. \\
            \hline
        \end{tabular}
    \end{table}

    \myred{To investigate whether WS-sTC retains the favorable exponential finite-size scaling of the original WS-TC method~\cite{Sundararaman13PRB}, we show in \cref{tab:C_tdl} the TDL-extrapolated HF exchange energies ($E_{\text{HFX}}^{\text{TDL}}$) of diamond obtained from pseudopotential calculations using WS-TC and WS-sTC with $\eta = 3$--$5$.
    The extrapolation is performed by fitting $E_{\text{HFX}}$ on three adjacent $k$-point meshes to the following exponential formula~\cite{Spence08PRB,Sundararaman13PRB}:
    \begin{equation}    \label{eq:tdl_exp}
        E_{\text{HFX}}(N_k)
            = E_{\text{HFX}}^{\text{TDL}} + A \exp\left(-B N_k^{1/3}\right).
    \end{equation}
    For comparison, we also include the TDL-extrapolated probe-charge Ewald results obtained via \cref{eq:tdl_gapped}.}

    \myred{For the WS-TC reference, the $(3^3, 4^3, 5^3)$ extrapolation agrees with the larger $(4^3, 5^3, 6^3)$ extrapolation to within $0.05$~$\text{m}E_{\text{h}}$ per atom, consistent with rapid exponential FSE decay and supporting the use of the latter as a TDL reference for FSE evaluation in \cref{tab:C_tdl}.
    WS-sTC with $\eta \geq 3$ reproduces the WS-TC reference with excellent accuracy: the $(3^3, 4^3, 5^3)$ extrapolation gives an FSE of $0.03$--$0.04$~$\text{m}E_{\text{h}}$ per atom, whereas the $(4^3, 5^3, 6^3)$ extrapolation agrees with WS-TC to the reported precision.
    By contrast, the probe-charge Ewald extrapolation based on polynomial FSE scaling leaves a sizable residual FSE of $0.75$~$\text{m}E_{\text{h}}$ per atom using $(3^3, 4^3, 5^3)$ and requires the larger $(5^3, 6^3, 7^3)$ mesh trio to reduce the FSE to $0.07$~$\text{m}E_{\text{h}}$ per atom.
    \Cref{tab:C_tdl} also shows the corresponding all-electron results for diamond, for which FSEs are measured relative to the WS-sTC result with $\eta = 5$ extrapolated from the largest available mesh trio, $(4^3,5^3,6^3)$.
    The resulting FSEs closely parallel those observed in the pseudopotential case, suggesting that WS-sTC also achieves exponential FSE scaling in all-electron calculations.
    Tables~S1--S7 present similar comparisons of TDL extrapolations for all other systems studied in this work and further support the rapid TDL convergence of WS-sTC in gapped systems.
    }

    \subsection{Computational cost}
    \label{subsec:computational_cost}

    \begin{figure}[!t]
        \centering
        \includegraphics[width=3in]{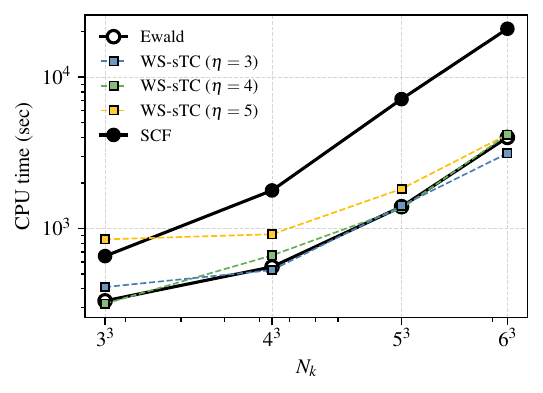}
        \caption{\myred{CPU time for RSDF initialization in all-electron HF calculations of diamond in the cc-pVDZ basis set using probe-charge Ewald and sTC with $\eta = 3$--$5$.
        For comparison, the CPU time for the full SCF calculation, excluding RSDF initialization, is also shown.
        All calculations were performed using 16 AMD EPYC 7763 CPU cores and a total of 64 GB of memory.}
        }
        \label{fig:C_timing}
    \end{figure}

    \myred{Finally, we use all-electron HF calculations of diamond in the cc-pVDZ basis set to assess the computational cost of RSDF-based sTC.
    The total cost of an RSDF-based HF calculation consists primarily of two parts: (i) the cost of RSDF initialization, which depends on the range-separation parameter $\omega$ and therefore varies with the sTC parameter $\eta$ according to \cref{eq:omega_Rin_cond}, and (ii) the cost of the SCF cycles, which is independent of $\omega$ and proportional to the number of cycles required for convergence.
    We find similar SCF convergence rates for the different FSE treatments, leading to nearly identical SCF costs.}

    \myred{\Cref{fig:C_timing} shows the CPU time for RSDF initialization using probe-charge Ewald and sTC with $\eta = 3$--$5$.
    The initialization cost of sTC with $\eta = 3$ and $4$ is comparable to that of probe-charge Ewald for all $k$-meshes tested.
    For $\eta = 5$, sTC noticeably increases the RSDF initialization cost on the $3^3$ and $4^3$ $k$-meshes, but yields similar CPU times on larger $k$-meshes.
    Moreover, the cost of the SCF cycles begins to dominate the total HF cost on larger $k$-meshes, making the total cost less sensitive to differences in RSDF initialization cost between FSE treatments.
    These results show that using sTC within RSDF introduces only a modest overhead relative to probe-charge Ewald, even when the smoothing parameter is set to $\eta = 5$.
    Therefore, in practical applications of sTC, there is considerable room to increase $\eta$ and improve accuracy without significantly increasing the computational cost.
    }

    \myred{We also note in passing that the cost of FFTDF-based sTC is determined by the size of the plane-wave basis needed to resolve the Gaussian pair densities, which is independent of $\eta$.
    For this reason, sTC and TC have identical cost within FFTDF.}

    \section{Conclusion}
    \label{sec:conclusion}

    In summary, we have introduced the sTC potential as a systematically improvable approximation to a parent TC potential for periodic HF exchange calculations.
    Gaussian smoothing of the truncation boundary, controlled by the dimensionless inverse relative smoothing width $\eta$, separates the sTC potential into short-range and long-range components that can be evaluated efficiently in real and reciprocal space.
    The resulting dual-space algorithm uses standard range-separated Gaussian integral kernels and applies to both pseudopotential and all-electron calculations with WS-cell truncation.
    Across insulating, semiconducting, layered, metallic, and molecular-crystal systems, sTC-based HF calculations with $\eta=3$ closely reproduce TC results for gapped systems, whereas $\eta=4$ is a more conservative choice for metals.
    These choices substantially accelerate convergence to the TDL relative to the probe-charge Ewald method while retaining accurate structural and spectral properties.

    In the future, the sTC approach can be extended in several directions.
    First, the isotropic Gaussian convolution used here is only one possible route for smoothing a parent TC potential.
    Anisotropy-aware Gaussian kernels and product-based smoothing functions~\cite{Ren21PRM,Zhang26JCTC} may provide greater efficiency for highly anisotropic truncation domains.
    Second, as in molecular density fitting, caching three-center integrals creates a storage bottleneck for RSDF in large systems.
    Integral-direct algorithms~\cite{Almlof82JCC,Haser89JCC,Bintrim22JCTC} that exploit locality~\cite{Manzer14JCTC,Wang20JCP} offer a promising route to bypass the storage bottleneck.
    Finally, the dual-space ERI algorithm is not restricted to periodic HF and may enable the use of (s)TC potentials in range-separated hybrid DFT~\cite{Lee22JCTC}, excited-state methods~\cite{Arruabarrena23PRB,Zhang26JCTC}, and periodic local-correlation theories~\cite{Ye24JCTC}.

    \section*{Supporting Information}

    See the Supporting Information for (i) crystal structures and basis-set data; (ii) direct sTC--TC differences in $E_{\text{HFX}}$ for diamond, silicon, h-BN, lithium, and aluminum; (iii) additional h-BN convergence results obtained using anisotropic $k$-point meshes; (iv) the dependence of the smoothing parameter $\omega$ on $k$-mesh size; (v) HF band structures of silicon; and (vi) TDL extrapolations of HF energies.

    \section*{Conflict of interest}
    The authors declare no competing interests.

    \section*{Data availability}

    The numerical data underlying the figures and tables in this study are available in Zenodo at \href{https://doi.org/10.5281/zenodo.22212380}{https://doi.org/10.5281/zenodo.22212380}.

    \section*{Acknowledgments}

    This work was supported by the National Science Foundation under Grant No.~CHE-2543461.
    H.Y.~thanks Min-Ye Zhang for helpful discussion.
    We acknowledge computing resources provided by the Division of Information Technology at the University of Maryland, College Park.

    \bibliography{refs}

\end{document}


\title{Supporting Information for: Smoothed truncated Coulomb potential for periodic Gaussian-basis Hartree--Fock exchange}

    \author{Gengzhi Yang}
    \affiliation{Joint Center for Quantum Information and Computer Science, University of Maryland, College Park, Maryland, 20742}
    \affiliation{Department of Mathematics, University of Maryland, College Park, Maryland, 20742}

    \author{Aamy Bakry}
    \affiliation{Department of Chemistry and Biochemistry, University of Maryland, College Park, Maryland, 20742}

    \author{Hong-Zhou Ye}
    \email{hzye@umd.edu}
    \affiliation{Department of Chemistry and Biochemistry, University of Maryland, College Park, Maryland, 20742}
    \affiliation{Institute for Physical Science and Technology, University of Maryland, College Park, Maryland, 20742}

    \maketitle

    \tableofcontents

    \clearpage

    \section{Geometries and basis-set data}

    Crystal structures for all systems considered in this work, provided in VASP POSCAR format, and the corresponding basis-set data files are available at

    \href{https://github.com/hongzhouye/supporting_data/tree/main/2026/sTC}{https://github.com/hongzhouye/supporting\_data/tree/main/2026/sTC}

    Diamond and silicon are represented by two-atom primitive cells of the diamond structure with lattice constants $a = 3.567~\text{\AA}$ and $5.431~\text{\AA}$, respectively.
    For h-BN, we use a four-atom primitive cell with $a = 2.501~\text{\AA}$ and $c = 6.351~\text{\AA}$.
    Lithium is represented by a two-atom body-centered-cubic conventional cell with $a = 3.510~\text{\AA}$, whereas aluminum is represented by a one-atom face-centered-cubic primitive cell with $a = 4.050~\text{\AA}$.
    The geometries of the \ce{CO2} and \ce{NH3} molecular crystals are taken from Ref.~\onlinecite{DellaPia24PRL}, and that of the tetrathiafulvalene (TTF) crystal is taken from Ref.~\onlinecite{Zhugayevych23JCTC}.

    \section{Additional convergence results}

    \begin{figure}[!h]
        \centering
        \includegraphics[width=5.5in]{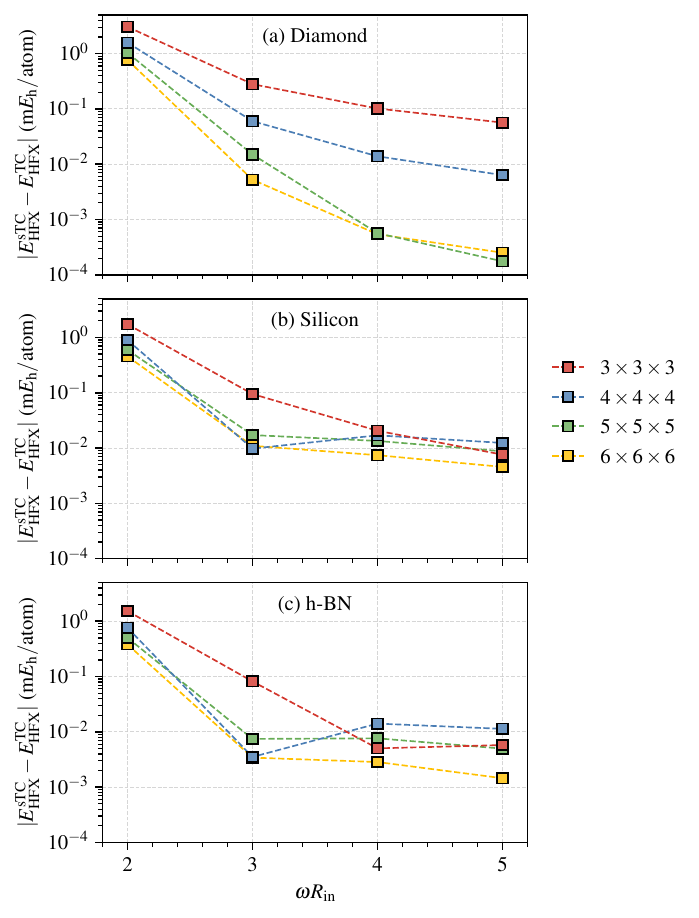}
        \caption{Absolute difference between the WS-sTC and WS-TC HF exchange energies per atom for diamond, silicon, and h-BN, calculated using GTH pseudopotentials.
        Results are shown as functions of $\eta=\omega R_{\mathrm{in}}$ for four $k$-point meshes.}
    \end{figure}

    \begin{figure}[!h]
        \centering
        \includegraphics[width=4.5in]{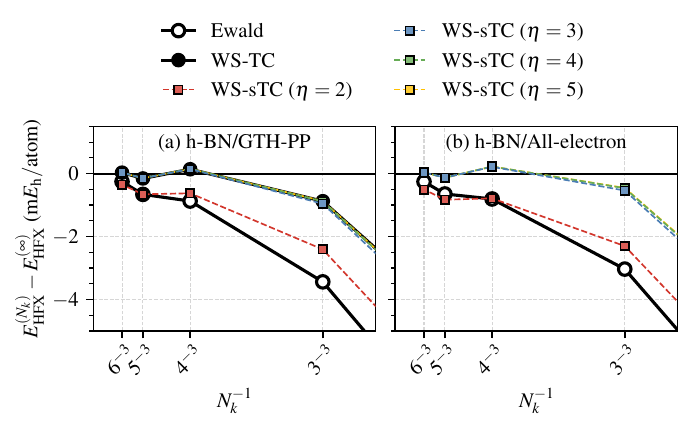}
        \caption{Finite-size error in the HF exchange energy per atom of h-BN obtained from (a) GTH-pseudopotential and (b) all-electron calculations.
        WS-sTC results with $\eta=2$--$5$ are compared with the probe-charge Ewald method; WS-TC is included for the pseudopotential calculations.
        These panels complement Fig.~2(c,f) of the main text by using isotropic $n \times n \times n$ $k$-point meshes for $n=3$--$6$.
        The resulting BvK supercells retain the unit-cell anisotropy, leading to irregular convergence toward the thermodynamic limit.
        Nonetheless, WS-TC gives smaller FSEs than probe-charge Ewald on the finite $k$-meshes shown.
        WS-sTC also retains its rapid convergence toward the WS-TC reference as $\eta$ increases, even for this anisotropic system.
        }
    \end{figure}

    \begin{figure}[!h]
        \centering
        \includegraphics[width=6.45in]{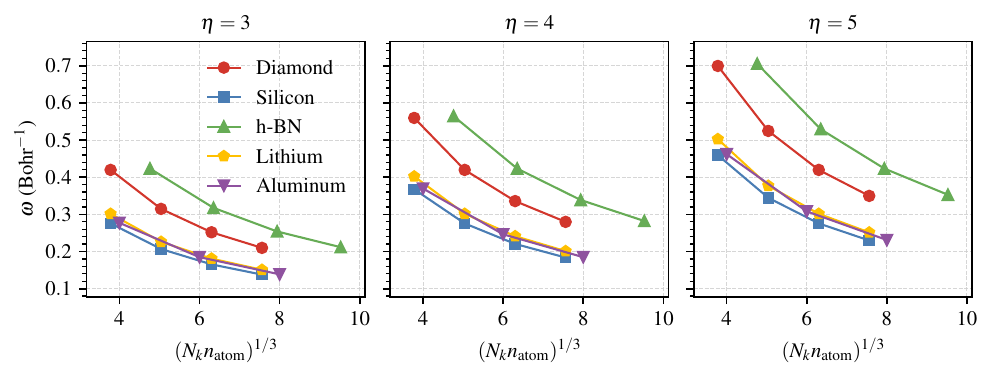}
        \caption{Dependence of the smoothing parameter $\omega$ on the linear size of the BvK supercell, $(N_k n_{\mathrm{atom}})^{1/3}$, for (a) $\eta=3$, (b) $\eta=4$, and (c) $\eta=5$.
        Here, $N_k$ is the number of sampled $k$-points and $n_{\mathrm{atom}}$ is the number of atoms in the unit cell.
        At fixed $\eta$, $\omega$ decreases as the BvK supercell is enlarged for all five solids.
        }
    \end{figure}

    \begin{figure}[!h]
        \centering
        \includegraphics[width=5in]{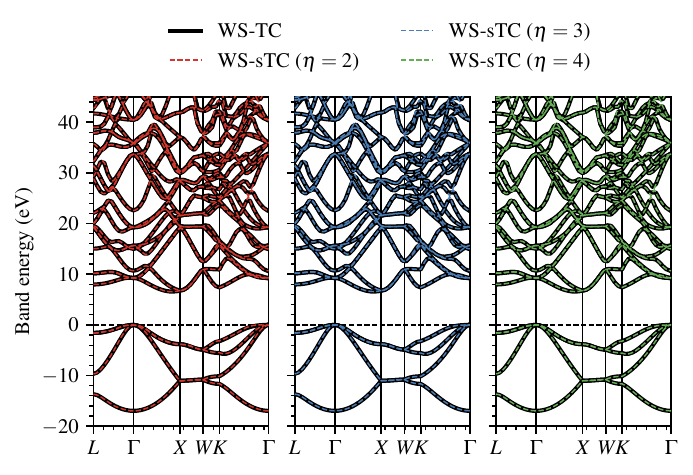}
        \caption{HF band structure of silicon calculated using WS-sTC with $\eta=2$--$4$ and WS-TC on a $4\times4\times4$ $k$-point mesh.
        The GTH pseudopotential and GTH-cc-pVTZ basis set are used.
        For each calculation, the band energies are shifted relative to its valence-band maximum.
        }
    \end{figure}

    \begin{figure}[!h]
        \centering
        \includegraphics[width=4.5in]{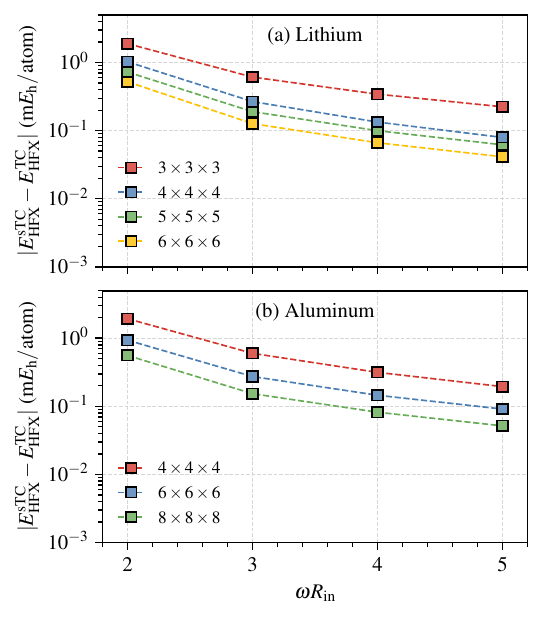}
        \caption{Absolute difference between the WS-sTC and WS-TC HF exchange energies per atom for (a) lithium and (b) aluminum, calculated using GTH pseudopotentials.
        Results are shown as functions of $\eta=\omega R_{\mathrm{in}}$ for four $k$-point meshes.}
    \end{figure}

    \clearpage

    \section{TDL extrapolation}

    \Cref{tab:Si_tdl,tab:hBN_tdl,tab:Li_tdl,tab:Al_tdl} compare the TDL-extrapolated per-atom HF exchange energies obtained with different FSE treatments and $k$-point meshes for Si, h-BN, Li, and Al, respectively.
    These tables follow the format of Table~III in the main text for diamond.
    For the gapped solids, probe-charge Ewald results are fitted to a quadratic polynomial in $1/N_k$, whereas WS-TC and WS-sTC results are fitted to an exponential in $N_k^{1/3}$, as described in the main text.
    For Li and Al, all methods are fitted to the metallic form $E_{\text{HFX}}(N_k) = E_{\text{HFX}}^{\text{TDL}} + A N_k^{-2/3} + B N_k^{-1}$.
    Taking the extrapolations from the largest available mesh trios for WS-TC in GTH-PP calculations and WS-sTC with $\eta = 5$ in all-electron calculations as references, we estimate the uncertainty of the largest three-point probe-charge Ewald extrapolation to be \textit{ca}.~$0.1$~$\text{m}E_{\text{h}}$ per atom for diamond, silicon, and aluminum, $0.3$~$\text{m}E_{\text{h}}$ per atom for lithium, and $\ll 0.1$~$\text{m}E_{\text{h}}$ per atom for h-BN.
    These small discrepancies support the use of the three-point probe-charge Ewald extrapolations as TDL references for Figs.~2 and 5 in the main text.

    \Cref{tab:CO2_tdl,tab:NH3_tdl,tab:TTF_tdl} present corresponding results for the three molecular crystals.
    Here, we report the HF total energy per molecule and use the same extrapolation formulas as for the gapped systems above.
    The estimated uncertainty of the largest three-point probe-charge Ewald extrapolation is well below $0.1$~kcal/mol per molecule for all three crystals, supporting its use as the TDL reference for Figs.~6 and 7 in the main text.

    \begin{table}[!h]
        \centering
        \caption{Same as Table~III in the main text, but for silicon.
        }
        \label{tab:Si_tdl}
        \begin{tabular}{llcccc}
            \hline\hline
            \multirow{2}*{Method} & \multirow{2}*{$k$-meshes}
                & \multicolumn{2}{c}{GTH-PP}
                & \multicolumn{2}{c}{All-electron}  \\
            \cmidrule(lr){3-4} \cmidrule(lr){5-6}
                &
                & $E_{\text{HFX}}^{\text{TDL}}/E_{\text{h}}$
                & FSE$/\text{m}E_{\text{h}}$
                & $E_{\text{HFX}}^{\text{TDL}}/E_{\text{h}}$
                & FSE$/\text{m}E_{\text{h}}$    \\
            \hline
            \multirow{3}{8em}{Ewald} & ($3^3, 4^3, 5^3$) & $-1.08151$ & $1.09$ & $-20.43355$ & $1.06$ \\
             & ($4^3, 5^3, 6^3$) & $-1.08217$ & $0.43$ & $-20.43420$ & $0.41$ \\
             & ($5^3, 6^3, 7^3$) & $-1.08249$ & $0.11$ & $-20.43451$ & $0.11$ \\
            \hline
            \multirow{2}{8em}{WS-TC} & ($3^3, 4^3, 5^3$) & $-1.08260$ & $0.00$ & \textemdash & \textemdash \\
             & ($4^3, 5^3, 6^3$) & $-1.08260$ & TDL ref. & \textemdash & \textemdash \\
            \hline
            \multirow{2}{8em}{WS-sTC ($\eta=3$)} & ($3^3, 4^3, 5^3$) & $-1.08260$ & $0.00$ & $-20.43463$ & $-0.02$ \\
             & ($4^3, 5^3, 6^3$) & $-1.08260$ & $0.00$ & $-20.43461$ & $0.00$ \\
            \hline
            \multirow{2}{8em}{WS-sTC ($\eta=4$)} & ($3^3, 4^3, 5^3$) & $-1.08260$ & $0.00$ & $-20.43463$ & $-0.02$ \\
             & ($4^3, 5^3, 6^3$) & $-1.08260$ & $0.00$ & $-20.43461$ & $0.00$ \\
            \hline
            \multirow{2}{8em}{WS-sTC ($\eta=5$)} & ($3^3, 4^3, 5^3$) & $-1.08260$ & $0.00$ & $-20.43463$ & $-0.02$ \\
             & ($4^3, 5^3, 6^3$) & $-1.08260$ & $0.00$ & $-20.43461$ & TDL ref. \\
            \hline
        \end{tabular}
    \end{table}

    \begin{table}[!h]
        \centering
        \caption{Same as Table~III in the main text, but for h-BN.
        Extrapolations use the anisotropic $3n \times 3n \times n$ $k$-mesh series ($n = 1, 2, 3$), which compensates for the unit-cell anisotropy.
        }
        \label{tab:hBN_tdl}
        \begin{tabular}{llcccc}
            \hline\hline
            \multirow{2}*{Method} & \multirow{2}*{$k$-meshes}
                & \multicolumn{2}{c}{GTH-PP}
                & \multicolumn{2}{c}{All-electron}  \\
            \cmidrule(lr){3-4} \cmidrule(lr){5-6}
                &
                & $E_{\text{HFX}}^{\text{TDL}}/E_{\text{h}}$
                & FSE$/\text{m}E_{\text{h}}$
                & $E_{\text{HFX}}^{\text{TDL}}/E_{\text{h}}$
                & FSE$/\text{m}E_{\text{h}}$    \\
            \hline
            \multirow{1}{8em}{Ewald} & ($9, 72, 243$) & $-1.66568$ & $-0.02$ & $-5.30430$ & $-0.02$ \\
            \hline
            \multirow{1}{8em}{WS-TC} & ($9, 72, 243$) & $-1.66566$ & TDL ref. & \textemdash & \textemdash \\
            \hline
            \multirow{1}{8em}{WS-sTC ($\eta=3$)} & ($9, 72, 243$) & $-1.66566$ & $0.00$ & $-5.30427$ & $0.00$ \\
            \hline
            \multirow{1}{8em}{WS-sTC ($\eta=4$)} & ($9, 72, 243$) & $-1.66566$ & $0.00$ & $-5.30427$ & $0.00$ \\
            \hline
            \multirow{1}{8em}{WS-sTC ($\eta=5$)} & ($9, 72, 243$) & $-1.66566$ & $0.00$ & $-5.30427$ & TDL ref. \\
            \hline
        \end{tabular}
    \end{table}

    \begin{table}[!h]
        \centering
        \caption{Same as Table~III in the main text, but for lithium.
        All methods use the metallic extrapolation formula described above.
        }
        \label{tab:Li_tdl}
        \begin{tabular}{llcccc}
            \hline\hline
            \multirow{2}*{Method} & \multirow{2}*{$k$-meshes}
                & \multicolumn{2}{c}{GTH-PP}
                & \multicolumn{2}{c}{All-electron}  \\
            \cmidrule(lr){3-4} \cmidrule(lr){5-6}
                &
                & $E_{\text{HFX}}^{\text{TDL}}/E_{\text{h}}$
                & FSE$/\text{m}E_{\text{h}}$
                & $E_{\text{HFX}}^{\text{TDL}}/E_{\text{h}}$
                & FSE$/\text{m}E_{\text{h}}$    \\
            \hline
            \multirow{2}{8em}{Ewald} & ($3^3, 4^3, 5^3$) & $-1.77815$ & $2.75$ & $-1.80362$ & $0.76$ \\
             & ($4^3, 5^3, 6^3$) & $-1.78057$ & $0.32$ & $-1.80425$ & $0.13$ \\
            \hline
            \multirow{2}{8em}{WS-TC} & ($3^3, 4^3, 5^3$) & $-1.77759$ & $3.30$ & \textemdash & \textemdash \\
             & ($4^3, 5^3, 6^3$) & $-1.78089$ & TDL ref. & \textemdash & \textemdash \\
            \hline
            \multirow{2}{8em}{WS-sTC ($\eta=3$)} & ($3^3, 4^3, 5^3$) & $-1.77785$ & $3.05$ & $-1.80360$ & $0.78$ \\
             & ($4^3, 5^3, 6^3$) & $-1.78079$ & $0.10$ & $-1.80435$ & $0.03$ \\
            \hline
            \multirow{2}{8em}{WS-sTC ($\eta=4$)} & ($3^3, 4^3, 5^3$) & $-1.77779$ & $3.11$ & $-1.80358$ & $0.80$ \\
             & ($4^3, 5^3, 6^3$) & $-1.78083$ & $0.07$ & $-1.80437$ & $0.01$ \\
            \hline
            \multirow{2}{8em}{WS-sTC ($\eta=5$)} & ($3^3, 4^3, 5^3$) & $-1.77774$ & $3.15$ & $-1.80357$ & $0.82$ \\
             & ($4^3, 5^3, 6^3$) & $-1.78085$ & $0.05$ & $-1.80438$ & TDL ref. \\
            \hline
        \end{tabular}
    \end{table}

    \begin{table}[!h]
        \centering
        \caption{Same as Table~III in the main text, but for aluminum.
        All methods use the metallic extrapolation formula described above.
        }
        \label{tab:Al_tdl}
        \begin{tabular}{llcccc}
            \hline\hline
            \multirow{2}*{Method} & \multirow{2}*{$k$-meshes}
                & \multicolumn{2}{c}{GTH-PP}
                & \multicolumn{2}{c}{All-electron}  \\
            \cmidrule(lr){3-4} \cmidrule(lr){5-6}
                &
                & $E_{\text{HFX}}^{\text{TDL}}/E_{\text{h}}$
                & FSE$/\text{m}E_{\text{h}}$
                & $E_{\text{HFX}}^{\text{TDL}}/E_{\text{h}}$
                & FSE$/\text{m}E_{\text{h}}$    \\
            \hline
            \multirow{2}{8em}{Ewald} & ($2^3, 4^3, 6^3$) & $-0.66720$ & $3.58$ & $-18.19823$ & $-2.65$ \\
             & ($4^3, 6^3, 8^3$) & $-0.67066$ & $0.12$ & $-18.19571$ & $-0.14$ \\
            \hline
            \multirow{2}{8em}{WS-TC} & ($2^3, 4^3, 6^3$) & $-0.66708$ & $3.70$ & \textemdash & \textemdash \\
             & ($4^3, 6^3, 8^3$) & $-0.67078$ & TDL ref. & \textemdash & \textemdash \\
            \hline
            \multirow{2}{8em}{WS-sTC ($\eta=3$)} & ($2^3, 4^3, 6^3$) & $-0.66714$ & $3.64$ & $-18.19834$ & $-2.77$ \\
             & ($4^3, 6^3, 8^3$) & $-0.67076$ & $0.02$ & $-18.19561$ & $-0.03$ \\
            \hline
            \multirow{2}{8em}{WS-sTC ($\eta=4$)} & ($2^3, 4^3, 6^3$) & $-0.66712$ & $3.66$ & $-18.19836$ & $-2.79$ \\
             & ($4^3, 6^3, 8^3$) & $-0.67077$ & $0.01$ & $-18.19558$ & $-0.01$ \\
            \hline
            \multirow{2}{8em}{WS-sTC ($\eta=5$)} & ($2^3, 4^3, 6^3$) & $-0.66712$ & $3.66$ & $-18.19837$ & $-2.80$ \\
             & ($4^3, 6^3, 8^3$) & $-0.67077$ & $0.01$ & $-18.19557$ & TDL ref. \\
            \hline
        \end{tabular}
    \end{table}

    \begin{table}[!h]
        \centering
        \caption{Same as Table~III in the main text, but for the HF total energy per molecule of the \ce{CO2} molecular crystal.
        Energies are reported in $E_{\text{h}}$ per molecule and FSEs in kcal/mol per molecule.
        }
        \label{tab:CO2_tdl}
        \begin{tabular}{llcc}
            \hline\hline
            \multirow{2}*{Method} & \multirow{2}*{$k$-meshes}
                & \multicolumn{2}{c}{All-electron}  \\
            \cmidrule(lr){3-4}
                &
                & $E_{\text{HF}}^{\text{TDL}}/E_{\text{h}}$
                & FSE$/$(kcal/mol)    \\
            \hline
            \multirow{2}{8em}{Ewald} & ($1^3, 2^3, 3^3$) & $-187.27076$ & $0.00$ \\
             & ($2^3, 3^3, 4^3$) & $-187.27077$ & $0.00$ \\
            \hline
            \multirow{2}{8em}{WS-sTC ($\eta=3$)} & ($1^3, 2^3, 3^3$) & $-187.27077$ & $0.00$ \\
             & ($2^3, 3^3, 4^3$) & $-187.27077$ & $0.00$ \\
            \hline
            \multirow{2}{8em}{WS-sTC ($\eta=4$)} & ($1^3, 2^3, 3^3$) & $-187.27076$ & $0.00$ \\
             & ($2^3, 3^3, 4^3$) & $-187.27076$ & $0.00$ \\
            \hline
            \multirow{2}{8em}{WS-sTC ($\eta=5$)} & ($1^3, 2^3, 3^3$) & $-187.27076$ & $0.00$ \\
             & ($2^3, 3^3, 4^3$) & $-187.27076$ & TDL ref. \\
            \hline
        \end{tabular}
    \end{table}

    \begin{table}[!h]
        \centering
        \caption{Same as Table~III in the main text, but for the HF total energy per molecule of the \ce{NH3} molecular crystal.
        Energies are reported in $E_{\text{h}}$ per molecule and FSEs in kcal/mol per molecule.
        }
        \label{tab:NH3_tdl}
        \begin{tabular}{llcc}
            \hline\hline
            \multirow{2}*{Method} & \multirow{2}*{$k$-meshes}
                & \multicolumn{2}{c}{All-electron}  \\
            \cmidrule(lr){3-4}
                &
                & $E_{\text{HF}}^{\text{TDL}}/E_{\text{h}}$
                & FSE$/$(kcal/mol)    \\
            \hline
            \multirow{2}{8em}{Ewald} & ($1^3, 2^3, 3^3$) & $-56.06715$ & $0.02$ \\
             & ($2^3, 3^3, 4^3$) & $-56.06717$ & $0.00$ \\
            \hline
            \multirow{2}{8em}{WS-sTC ($\eta=3$)} & ($1^3, 2^3, 3^3$) & $-56.06720$ & $-0.02$ \\
             & ($2^3, 3^3, 4^3$) & $-56.06717$ & $0.00$ \\
            \hline
            \multirow{2}{8em}{WS-sTC ($\eta=4$)} & ($1^3, 2^3, 3^3$) & $-56.06717$ & $0.00$ \\
             & ($2^3, 3^3, 4^3$) & $-56.06717$ & $0.00$ \\
            \hline
            \multirow{2}{8em}{WS-sTC ($\eta=5$)} & ($1^3, 2^3, 3^3$) & $-56.06717$ & $0.00$ \\
             & ($2^3, 3^3, 4^3$) & $-56.06717$ & TDL ref. \\
            \hline
        \end{tabular}
    \end{table}

    \begin{table}[!h]
        \centering
        \caption{Same as Table~III in the main text, but for the HF total energy per molecule of the TTF molecular crystal.
        Energies are reported in $E_{\text{h}}$ per molecule and FSEs in kcal/mol per molecule.
        Extrapolations use the anisotropic $n \times 2n \times n$ $k$-mesh series ($n = 1, 2, 3$), which compensates for the unit-cell anisotropy.
        }
        \label{tab:TTF_tdl}
        \begin{tabular}{llcc}
            \hline\hline
            \multirow{2}*{Method} & \multirow{2}*{$k$-meshes}
                & \multicolumn{2}{c}{All-electron}  \\
            \cmidrule(lr){3-4}
                &
                & $E_{\text{HF}}^{\text{TDL}}/E_{\text{h}}$
                & FSE$/$(kcal/mol)    \\
            \hline
            \multirow{1}{8em}{Ewald} & ($2, 16, 54$) & $-1818.80812$ & $0.05$ \\
            \hline
            \multirow{1}{8em}{WS-sTC ($\eta=3$)} & ($2, 16, 54$) & $-1818.80818$ & $0.01$ \\
            \hline
            \multirow{1}{8em}{WS-sTC ($\eta=4$)} & ($2, 16, 54$) & $-1818.80816$ & $0.02$ \\
            \hline
            \multirow{1}{8em}{WS-sTC ($\eta=5$)} & ($2, 16, 54$) & $-1818.80820$ & TDL ref. \\
            \hline
        \end{tabular}
    \end{table}

    \clearpage

    \bibliography{refs_si}